\documentclass[reprint,aps,pre]{revtex4-1}

\usepackage{amsmath,amsfonts,amscd,amssymb,graphicx,subeqnar}
\usepackage[bbgreekl]{mathbbol}

\DeclareMathAlphabet{\mathpzc}{OT1}{pzc}{m}{it}
 
\usepackage{dcolumn}
\usepackage{overpic}
\usepackage{color}
\usepackage{adjustbox}
\usepackage{float}
\usepackage{algorithm}
\usepackage{algpseudocode}
\algnewcommand\algorithmicinput{\textbf{Input:}}
\algnewcommand\Input{\item[\algorithmicinput]}

\usepackage{nomencl} 
\usepackage{etoolbox}
\makenomenclature

\DeclareSymbolFontAlphabet{\mathbb}{AMSb}
\DeclareSymbolFontAlphabet{\mathbbl}{bbold}

\renewcommand{\nomgroup}[1]{%
  \ifstrequal{#1}{L}{\item[\textbf{Latin Symbols}]}{%
  \ifstrequal{#1}{G}{\item[\textbf{Greek Letters}]}{%
  \ifstrequal{#1}{A}{\item[\textbf{Acronyms}]}{}}}}

\renewcommand{\vec}[1]{\mathbf{#1}}

\newcommand{\norma}[1]{\left|\left|{#1}\right|\right|}

\usepackage{hyperref}

\makeatletter
\@ifundefined{sf@counterlist}{\def\sf@counterlist{}}{}
\makeatother

\usepackage[hyperref,dvipsnames]{xcolor}

\definecolor{bluePolimi}{RGB}{22, 44, 80}
\definecolor{lightBluePolimi}{RGB}{91, 122, 172}
\definecolor{redPolimi}{RGB}{180, 0, 0}
\definecolor{greenPolimi}{RGB}{78, 172, 91}
\definecolor{green2}{RGB}{0, 110, 0}

\makeatletter
\hypersetup{colorlinks=true}
\AtBeginDocument{\@ifpackageloaded{hyperref}
  {\def\@linkcolor{blue}
   \def\@anchorcolor{red}
   \def\@citecolor{red}
   \def\@filecolor{red}
   \def\@urlcolor{redPolimi}
   \def\@menucolor{red}
   \def\@pagecolor{cyan}
\begingroup
  \@makeother\`%
  \@makeother\=%
  \edef\x{%
    \edef\noexpand\x{%
      \endgroup
      \noexpand\toks@{%
        \catcode 96=\noexpand\the\catcode`\noexpand\`\relax
        \catcode 61=\noexpand\the\catcode`\noexpand\=\relax
      }%
    }%
    \noexpand\x
  }%
\x
\@makeother\`
\@makeother\=
}{}}
\makeatother

\usepackage{comment}
\usepackage{tabularx}
\usepackage{booktabs} 

\begin{document}

\title{Robust State Estimation from Partial Out-Core Measurements with Shallow Recurrent Decoder for Nuclear Reactors}

\author{Stefano Riva$^{*}$, Carolina Introini$^{*}$, Antonio Cammi$^{*}$, J. Nathan Kutz$^{\dag}$}
\affiliation{$^{*}$Politecnico di Milano, Dept. of Energy, CeSNEF - Nuclear Engineering Division, 20156 Milan, Italy}
\affiliation{$^\dag$Department of Applied Mathematics and  Electrical and Computer Engineering, University of Washington, Seattle, WA 98195}

\begin{abstract} 
Reliable, real-time state estimation in nuclear reactors is of critical importance for monitoring, control and safety.  It further empowers the development of digital twins that are sufficiently accurate for real-world deployment. As nuclear engineering systems are typically characterised by extreme environments, their in-core sensing is a challenging task, even more so in Generation-IV reactor concepts, which feature molten salt or liquid metals as thermal carriers. The emergence of data-driven methods allows for new techniques for accurate and robust estimation of the full state space vector characterising the reactor (mainly composed by neutron fluxes and the thermal-hydraulics fields). These techniques can combine different sources of information, including computational proxy models and local noisy measurements on the system, in order to robustly estimate the state.  This work leverages the {\em Shallow Recurrent Decoder} (SHRED) architecture to estimate the entire state vector of a reactor from three, out-of-core time-series neutron flux measurements alone.  Specifically, the Molten Salt Fast Reactor, in the EVOL geometry (Evaluation and Viability of Liquid Fuel Fast Reactor System project), is demonstrated as a test case, with neutron flux measurements alone allowing for reconstruction of the 20 coupled field variables of the dynamics. This approach can further quantify the uncertainty associated with the state estimation due to its considerably low training cost on compressed data. The accurate reconstruction of every characteristic field in real-time makes this approach suitable for monitoring and control purposes in the framework of a reactor digital twin.
\end{abstract}

\maketitle


\section{Introduction}

To achieve the goals set by the COP28, energy transition must take a central role. In a world that sees an increasing demand for clean and sustainable energy, \textit{new nuclear} technologies present a crucial and complementary solution to renewable energy sources. Nuclear power is a dependable energy source, with high availability and fixed costs. As such, nuclear energy has, jointly with renewable energy, a major role to play in the clean energy transition. Innovative reactor concepts are currently being designed by considering their possible integration with other sources of energy and by following four core tenets as identified by the Generation-IV International Forum \cite{GenIV-RoadMap}: (i) efficient use of the fuel, (ii) reduced waste production, (iii) economical competitiveness, and (iv) meeting stringent standards of safety and proliferation resistance.

The development of innovative reactor concepts poses several challenges in terms of design, monitoring and safety. For example, these reactors could, in principle, work with spent fuel from traditional reactors; two out of the six proposed Generation-IV concepts operate with liquid metals as thermal carriers; whereas, one of them, the Molten Salt Fast Reactor (MSFR), adopts a liquid molten salt fuel homogeneously mixed with the coolant \cite{brovchenko2013optimization}. Moreover, they are usually designed to work in a fast neutron spectrum \cite{DuderstadtHamilton}. This condition makes the in-core environment more hostile due to the higher radiation field, with greater damage (and therefore lower useful lifetime) to structural materials and to in-core sensors. To address the challenge of modeling emerging reactor concepts, we leverage a {\em Shallow Recurrent Decoder} (SHRED) architecture to robustly map out-of-core measurements to full state estimates in the reactor, even fields unmeasured by the sensors. This provides a viable pathway for the construction of a digital twin capable of accurate monitoring of the fission process for safety critical applications.

The production of energy from nuclear reactors has always been subject to rigorous safety criteria at both design and operation levels. This topic has become even more relevant as nuclear reactors are planned to be included in hybrid energy grids along with renewable sources \cite{international2023iaea, RUTH2014684}.  The intermittent nature of these renewable sources require the nuclear reactor to reliably operate in load-following mode, which requires being able to quickly assess the new state of the system following a rapid change in power, whilst ensuring safety constraints at all times. This requires having access in real-time to all the dominant quantities of interest describing the physical behaviour of the system (e.g., temperature, neutron flux, power density and coolant velocity). Fast, accurate and reliable \textit{digital twin} (DT) models of the physical plant \cite{Grieves_DT} are consequently of growing interest in the nuclear engineering community \cite{mohanty_physics-infused_2021, mohanty_development_2021, gong_data-enabled_2022}.

A key objective associated with DTs consists of developing an inverse model capable of inferring the full state (characterised by all the different fields of interest) from point (sensor) measurements \cite{argaud_sensor_2018, gong_parameter_2023, gong_efficient_2022}. These inverse models can be constructed by combining mathematical modelling and experimental information to provide a full state estimation in a quick and reliable way. These approaches are critical for real-time control, monitoring of the nuclear reactor, and ensuring safety standards.  Furthermore, fast models allow for performing sensitivity analysis and uncertainty quantification in reasonable computational times. Indeed, having a reliable tool that provides state estimation, identifies the onset of potential accidental scenarios, and characterizes the unexpected behaviour of experimentally unobservable quantities is essential for the development of efficient and "autonomous" control systems. In fact, as mentioned above, nuclear reactor cores are characterised by a harsh and hostile environment, which makes in-core sensing a non-trivial task. This challenge holds
especially true in the MSFR, which, due to the liquid nature of the core, requires the local measurements to be collected from sensors placed outside the core \cite{ICAPP_plus2023, PHYSOR24_MSFR_outcore}. Therefore, a DT of a nuclear reactor must be able to provide in real-time a complete estimation of the whole state of the reactor, integrating together the information of models and measurements while also being able to identify and predict accidental scenarios.

The simulation of mathematical models typically requires the numerical solution of parameter-dependent Partial Differential Equations (PDEs), which is not feasible for real-time applications \cite{Kapteyn_Willcox_2020}. Only the recent developments in Reduced Order Modelling (ROM) approaches \cite{lassila_model_2014, rozza_model_2020} have opened new possibilities towards reliable and efficient DTs for engineering systems. These techniques aim to produce a reduced representation of the solution manifold and provide an approximation that is both sufficiently accurate and obtained with reasonably low computational costs. Among ROM techniques, non-intrusive methods are more suited for the combination of models and measurements. Moreover, with respect to intrusive approaches \cite{quarteroni2015reduced}, they do not require the knowledge of the governing PDEs. They have been widely used in the literature to combine models with sensors with notable examples including the Gappy Proper Orthogonal Decomposition \cite{brunton_data-driven_2022}, Generalised Empirical Interpolation Method \cite{maday_generalized_2015} or the Parameterised-Background Data-Weak formulation \cite{maday_parameterized-background_2014}. 

A broader class of data-driven models have also emerged as the leading paradigm for learning input-output maps \cite{brunton_data-driven_2022}, including mapping temporal trajectories of measurements to the state space \cite{williams2022data}, making them applicable for state estimation purposes in nuclear reactors \cite{gong_data-enabled_2022, gong_efficient_2022}. Indeed, the non-intrusive techniques mentioned above include a step which determines the optimal sensor placement to obtain a well-conditioned inverse problem upon which state estimates rely \cite{binev_greedy_2018, Manohar2018}. Within the nuclear engineering framework, a two-step approach has been proposed by \cite{INTROINI2023109538} and \cite{gong_parameter_2023}: the characteristic parameter, producing some local measurements, is estimated with an inverse optimisation problem and then used to calculate the temporal and parametric trajectories embedded in the reduced coefficients. Even though promising results have been obtained even using out-core measurements \cite{ICAPP_plus2023}, optimisation problems are often computationally expensive even when solved in a reduced coordinate system; in fact, this was the main computational bottleneck of the work done by the authors in \cite{INTROINI2023109538}.


The {\em SHallow REcurrent Decoder} is a neural network architecture that, following a training and learning process, maps the trajectories of sensor measurements to a latent space, thus encoding the dynamics of the high-dimensional space. In its basic principles, SHRED can be seen as a generalization of the separation of variables method for solving PDEs~\cite{williams2022data, ebers2023leveraging, kutz_shallow_2024}.  
This work applies the SHRED architecture to the Molten Salt Fast Reactor considering the EVOL geometry \cite{brovchenko2013optimization}: in particular, the overall state vector is to be reconstructed using out-core measurements a single field, following \cite{ICAPP_plus2023, PHYSOR24_MSFR_outcore}. The system state includes the energy fluxes, the neutron precursors \cite{DuderstadtHamilton}, the decay heat groups \cite{aufiero2014development} and the thermal-hydraulics fields, i.e., the pressure, the temperature and the velocity; in this work, only one of the energy fluxes is supposed to be observable (this choice is arbitrary and it may depend on the data availability on the plant site), as in general, fluxes measures are easier with respect to others, like the precursors concentration or the velocity. Moreover, the sensors are only placed in the external solid reflector region to avoid in-core placement, an almost impossible task due to the fluid nature of the MSFR. The SHRED architecture is such that the training phase is quite rapid and inexpensive and the required hyperparameter tuning on the net parameters is minimal \cite{kutz_shallow_2024}, to the point that it can be easily performed on personal computers \cite{williams2022data}. These features make this architecture well-suited for different physical problems and noisy data: more than one SHRED model can be trained quickly so that different predictions of the temporal dynamics can be assembled to obtain a robust reconstruction, with the additional benefit of providing a mean value of the temporal dynamics and an uncertainty quantification of the estimation itself.

The paper is structured as follows: firstly, in Section \ref{sec: shred}, the basics of the SHRED architecture are discussed; Section \ref{sec: msfr} describes the Molten Salt Fast Reactor and the numerical modelling related to the full-order model (i.e., the source of training data); the numerical results are then presented in Section \ref{sec: num-res}; lastly, the main conclusion are drawn in Section \ref{sec: conclusions}.

\begin{figure*}[t]
    \centering
    \includegraphics[width=1\linewidth]{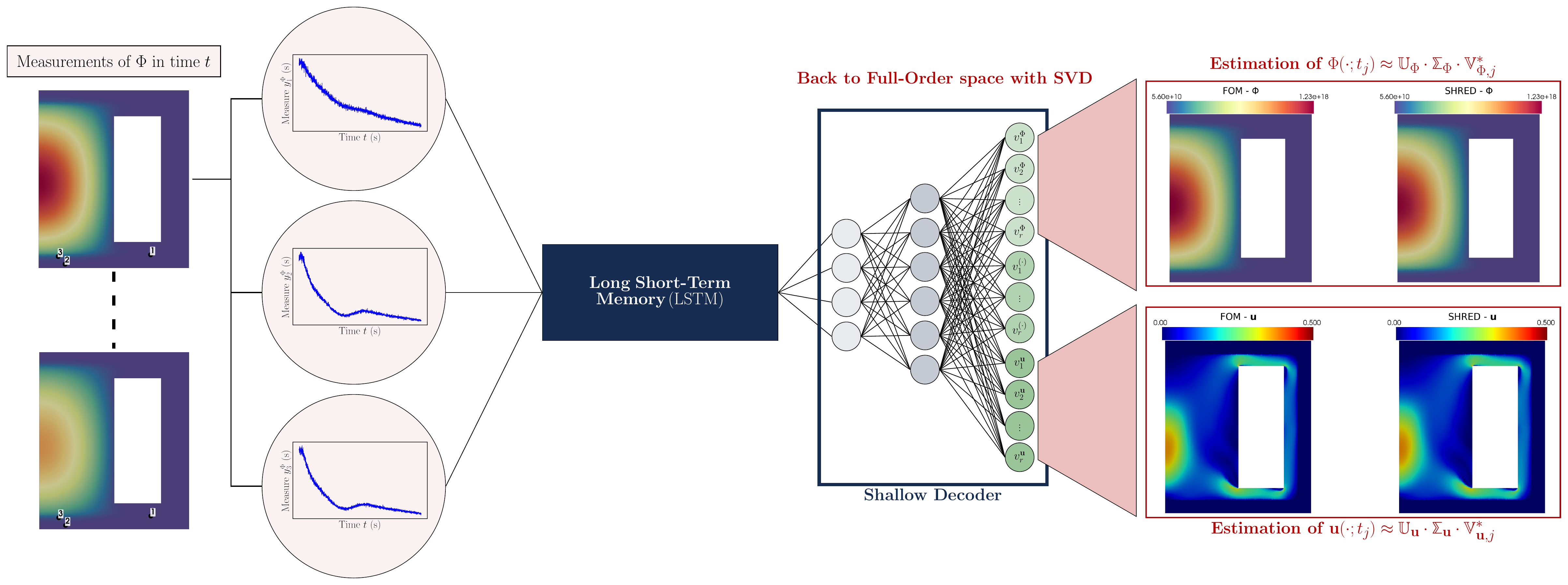}
    \caption{SHRED architecture applied to Nuclear Reactors.  Three out-of-core sensors are used to measure a single field variable.  The sensor time series is used to construct a latent temporal sequence model which is mapped to compressive representations of all spatio-temporal field variables.  The compressive representations can then be mapped to the original state space by the singular value decomposition.  The compressive representation allows for laptop level training in minutes even on 20 high-dimensional field variables.}
    \label{fig: shred}
\end{figure*}

\section{SHallow REcurrent Decoder}\label{sec: shred}

The SHRED \cite{williams2022data, ebers2023leveraging, kutz_shallow_2024} architecture is a novel data-driven technique based on the combination of a Long Short-Term Memory (LSTM) \cite{hochreiter1997long}, a Shallow Decoder Network (SDN) \cite{erichson2020shallow} and the compression given by the Singular Value Decomposition (SVD), also known as Proper Orthogonal Decomposition \cite{brunton_data-driven_2022, rozza_model_2020}. The performance of the SHRED architecture was proven in \cite{williams2022data}, where given the same physical problems this structure outperformed, in terms of reconstruction accuracy, state-of-the-art linear and non-linear techniques given as low as three sensors. Even more promising, SHRED selects the sensors randomly among the available locations, whereas other non-intrusive methods typically adopt hierarchical greedy algorithms to identify the optimal position of the sensors, a step in the offline phase that often represents one of the computational bottlenecks. Instead, SHRED incorporates the temporal trajectory of the measurements through the LSTM to learn the evolution of the temporal dynamics embedded in the latent space generated by the SVD with the SDN.

Figure~\ref{fig: shred} highlights the architecture of SHRED. As a first step, the neural network is trained to map the measures to a compressed space, such as the low-rank approximation given by the SVD \cite{brunton_data-driven_2022}. SHRED maps only a few measurements $\vec{y}$ of a single observable field to the latent dynamics given by the reduced coefficients $v_n(t)$ of each characteristic quantity of interest (for the specific case of this paper, these quantities are those that describe the physics of the nuclear reactor, such as neutron fluxes, temperature and velocity). This first compression with SVD approximates the full-order space of the snapshots to a low-rank finite dimensional space spanned by the first $r$-rank left singular eigenvectors of the SVD, allowing for a significant reduction in the training cost to the point that the SHRED model can be trained in minutes even on a personal laptop \cite{kutz_shallow_2024}. Due to this, it becomes cheap to train different SHRED models for different sensor configurations, i.e., for other random sensor locations, to obtain more than one estimation of the state such that a measure of the uncertainty associated with the prediction can be provided. This way of proceeding can be useful to generate state estimations which are robust against noisy measurements collected on the physical system itself; furthermore, it can be extended to identify possible malfunctions to the sensors, since more than one estimation is given. 

In the SHRED architecture, the recurrent neural network~\cite{lipton2015critical} is an LSTM~\cite{hochreiter1997long,yu2019review} network, which models a time sequence of measurements, or trajectory, from a limited number of point sensors in any one of the temporal fields of the nuclear reactor dynamics; the LSTM itself constructs a latent space representation of the dynamics given a time-lagged embedding, which has been shown to be related to Takens embedding theory~\cite{takens1981lnm}. The latent space then projects the data through an SDN~\cite{erichson2020shallow} back to the latent space of the fields of interest, to be later de-compressed into the full-order space with the SVD.  

As already mentioned, SHRED has a theoretical basis in the PDE theory of separation of variables \cite{kutz_shallow_2024} along with the Takens embedding theorem~\cite{takens1981lnm}.  As already detailed in Williams et al  \cite{kutz_shallow_2024}, SHRED exploits the fact that time measurements are equivalent to spatial measurements in a spatio-temporal system. Thus sensor trajectory histories can completely characterize the dynamics regardless of measurement location (unless a region is statistically independent of other regions).  Moreover, using the fact that $N$ first order PDEs can be rewritten as an $N$th order PDE, it can be shown that a single field variable encodes all other variables coupled to it through the evolution of its time dynamics.  This is the basis of the Takens embedding theorem that guarantees that sensor trajectory information from a single field guarantees a diffeomorphic representation of all other fields.  With training, SHRED uniquely determines the diffeomorphism to be fixed to the original spatio-temporal fields. 
In this sense, the SHRED architecture can be conceived as a generalization to nonlinear PDEs, whereas a rigorous justification is available for linear PDEs only. As already shown in \cite{williams2022data, ebers2023leveraging, kutz_shallow_2024}, SHRED provides important advantages with respect to standard data-driven ROM methods including:
\begin{itemize}
    \item the ability to use only three sensors (even randomly selected) for reconstructing the entire dynamics (all fields) of a physical system;
    \item the ability to train on compressed data, spanned by the SVD basis and thus enabling laptop level training in minutes;
    \item the ability to measure a single field variable (the most convenient) and reconstruct coupled spatio-temporal fields that are not observable;
    \item minimal hyper-parameter tuning; in fact, the same architecture for the neural network can be adopted for different physical problems without the need to optimise the number of neurons and layers.
\end{itemize}

As noted, the SHRED architecture is agnostic toward sensor placement \cite{williams2022data, kutz_shallow_2024}; namely, there is no need to determine the optimal sensor configuration \cite{argaud_sensor_2018}, setting itself apart from most data-driven methods in which selecting the optimal positions for the sensors (that is, the ones that allow retrieving the most information) is a key part of the training phase, especially for safety-critical industries such as the nuclear one
 \cite{ICAPP_plus2023, gong_reactor_2024}. Furthermore, architectures like SHRED can provide full-state reconstruction with even out-core measurements of a single quantity that might be the simplest (least expensive) to diagnose. In harsh environments like the Molten Salt Fast Reactor, this is crucial for monitoring and safety, especially in the long term.

The SHRED architecture has been implemented in Python using the PyTorch package \cite{pytorch} and adopting the code developed by \cite{williams2022data, kutz_shallow_2024}. Both the  LSTM and the SDN network are composed by 2 hidden layers, with the number of neurons reported in Table \ref{tab: SHRED-architecture}.
\begin{table}[t]
    \centering
    \begin{tabular}{c|c|c}
        \toprule $\,$ & Hidden Layer 1 & Hidden Layer 2 \\ \midrule
        LSTM & 64 & 64 \\ \midrule
        SDN & 350 & 400  \\ \bottomrule
    \end{tabular}
    \caption{Architecture of the SHRED networks.}
    \label{tab: SHRED-architecture}
\end{table}

\section{Molten Salt Fast Reactor}\label{sec: msfr}

Among the innovative Generation-IV \cite{GenIV-RoadMap} reactors, the Molten Salt Fast Reactor has the unique feature of a liquid Thorium-based fuel homogeneously mixed with the thermal carrier; in particular, the fuel is a mixture of $^7$LiF and ThF$_4$ (77.5 – 22.5 mol\%), having a melting temperature of 838 K. Given a total fuel inventory of 9 m$^3$, this reactor can produce up to 3 GW$_{\text{th}}$ thermal \cite{serp_molten_2014}. The liquid nature of the MSFR fuel does not allow in-core sensing, as usually performed in water-cooled reactors, thus restricting measurements to be collected with out-core sensors; additionally, the fast energy spectrum of the reactor \cite{brovchenko2013optimization} makes the core environment very hostile \cite{ICAPP_plus2023}. Accordingly, the sensor placement will not be performed randomly on the whole domain, but the (random) choice is restricted to the reflector region only, as in \cite{PHYSOR24_MSFR_outcore}.

\begin{figure}[t]
    \centering
    \hspace*{-.2in}
    \includegraphics[width=1.05\linewidth]{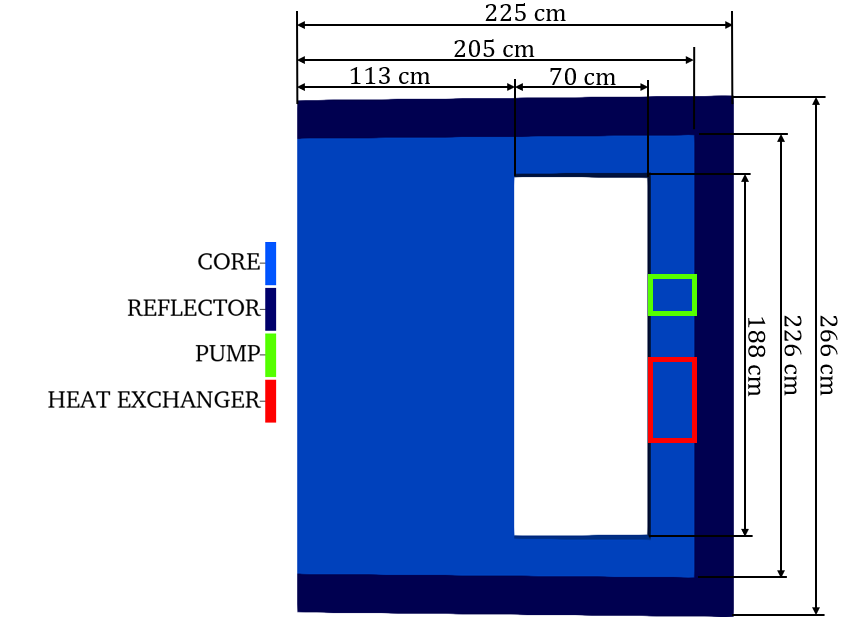}
    \caption{OpenFOAM simulation domain with the main dimensions and the primary loop components. The geometry refers to a 2D axisymmetric wedge of the EVOL geometry of the European MSFR design, including the shown components (molten salt fuel in light blue, Hastelloy reflector in dark blue, primary pump in green, heat exchanger with the intermediate cycle in red). The blank hole represents the solid salt fertile blanket, not simulated in the present model.}
    \label{fig: evol-geom}
\end{figure}

As a test case for the SHRED architecture, this work uses the 2D axisymmetric wedge (5°) of the EVOL geometry of the MSFR \cite{brovchenko2013optimization}, with an important modification: indeed, the present test case includes an external boundary of thickness 20 cm to mimic the presence of the Hastelloy reflector \cite{PHYSOR24_MSFR_outcore}: this subdomain is the only region in which, during the training phase, sensor locations can be randomly sampled. The thermo-physical properties of this region are those of nickel-based alloys, with the reflector neutronic group constants computed using the Monte Carlo code OpenMC \cite{OpenMC}, whereas data for the fuel were taken from \cite{aufiero2014development}. 
Thus, the simulation domain $\Omega$ includes two regions with different properties: the liquid core $\Omega_{\text{core}}$ and the solid reflector $\Omega_{\text{refl}}$. Figure \ref{fig: evol-geom} depicts the simulated domain along with its main dimensions, including the primary loop components (pump and heat exchanger). The white cavity represents the location of the fertile blanket, not modelled in the present work. 

The adopted numerical solver, developed at Politecnico di Milano, allows performing coupled neutronics and thermal-hydraulics simulations within the OpenFOAM environment \cite{aufiero2014development}. More in detail, the thermal-hydraulic sub-solver implements the incompressible single-phase version of the Reynolds-Averaged Navier-Stokes (RANS) equations with the Realizable $\kappa-\varepsilon$ turbulence model and the Boussinesq approximation to consider buoyancy effects (under the assumption that thermo-physical properties are constant within the same region); the neutronic sub-solver adopts the multi-group neutron diffusion approximation and includes transport equations for the delayed neutrons and the decay heat precursors. The Doppler and thermal expansion effects for the neutronic feedback coefficients have been modelled using, respectively, a linear and a logarithmic term correcting the reference group constants; furthermore, a momentum source and a heat sink represent the primary loop pump and the heat exchanger, respectively. For the interested reader, a summary of the governing equations, discretised using the Finite Volume Method in OpenFOAM, is reported in Appendix \ref{app: msfr-gov-eqn}. The mesh consists of 46424 hexahedral and 266 prismatic cells.

\begin{figure}[t]
    \centering
    \includegraphics[width=1\linewidth]{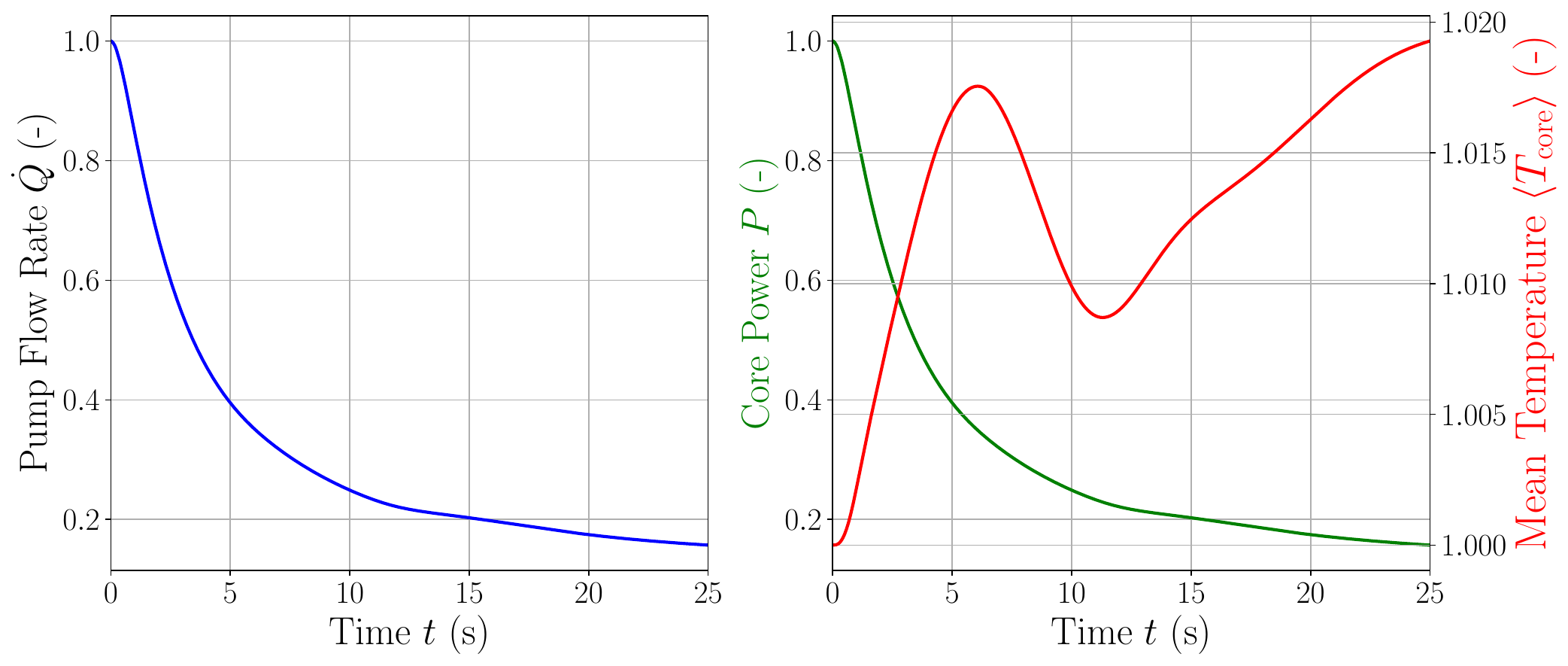}
    \caption{Evolution in time of global core quantities, each normalised with respect to the initial condition, following an unprotected failure of the primary pump and an exponential decrease of the pump flow rate, up to 25 seconds from the initiating event. On the left, the pump flow rate (input), whereas on the right the core power and the average temperature (outputs).}
    \label{fig: evol-core-qties}
\end{figure}

The transient considered in this work to generate the training dataset for SHRED is the accidental scenario of symmetric and unprotected failure of the primary pump, namely an Unprotected Loss of Fuel Flow (ULOFF). In this scenario, the flow rate of the pump is decreased exponentially (Figure \ref{fig: evol-core-qties}) resulting in a consequent decrease of the velocity magnitude in the reactor. In the first few seconds of the simulation, the reduction in flow rate translates into an increase in the power-to-flow ratio, increasing the average temperature, since there is a strong change in the dynamics from a forced convection problem to a natural circulation one. As the temperature rises in the core, the thermal feedback coefficients of the neutronics \cite{DuderstadtHamilton} come into play, making the overall power decrease exponentially (Figure \ref{fig: evol-core-qties}). The selected time interval of the simulation of the ULOFF transient is 25 seconds, with a saving time of $\Delta t = 0.05 $s resulting in $N_t= 500$ snapshots. The elapsed CPU time for simulating this accidental scenario is around 10 hours. 

The MSFR is a complex nuclear reactor, characterised by several fields that describe the neutron economy and the thermal-hydraulics: in particular, for this case, six group flux in energy $\{\phi_g\}_{g=1}^6$, eight groups of delayed neutrons $\{c_k\}_{k=1}^8$ and three decay heat groups $\{d_i\}_{i=1}^3$ are considered for the neutronic side, to which the thermal-hydraulics triplet, namely pressure, temperature and velocity $(p, T, \vec{u})$, must be added. Except for the velocity $\vec{u}$, all the others are scalar fields. Overall the full-order state space vector $\mathcal{V}$ is represented by 20 different fields, i.e. 
\begin{equation}
    \mathcal{V} = \left[\phi_1, \dots, \phi_6, c_1, \dots, c_8, d_1, \dots, d_3, p, T, \vec{u}\right] 
\end{equation}

In real engineering systems, it is not always possible to have access to all the quantities of interest, such as those reported above; however, for strongly coupled problems as the MSFR test case, which are inherently multi-physics in nature and for which separation of the various physics is impossible, each field carries some information about other quantities, thus, information on the un-observable fields can be extracted from the observable ones \cite{gong_data-enabled_2022, INTROINI2023109538}: for instance, in the present test case temperature and velocity are connected through the buoyancy and the energy balance, and the neutron fluxes strongly depend on the temperature distribution \cite{aufiero2014development}. It is then legitimate to investigate the indirect reconstruction problem for such cases, assuming only few fields (as low as one) can be directly measured.

\subsection{SHRED Setup for Multi-Physics systems}

The SHRED architecture learns a map between three measurements of an observable field and the reduced representation through the SVD of the full-order state space vector; therefore, it is necessary to obtain time-series data of an observable field and perform the SVD of each characteristic field.

Starting from the latter, each field (generically indicated as $\psi$) is organized in the form of a snapshot matrix $\mathbb{X}_\psi\in\mathbb{R}^{\mathcal{N}_h\times N_t}$ such that the $j-$th column represents a snapshot at a given time $t_j$, given $\mathcal{N}_h$ the dimension of the spatial mesh. The snapshots are normalised to the mean value at the initial time to obtain variables within the same range (as the different fields of the MSFR are characterised by vastly different time scales: as an example, the neutron flux reaches $\sim 10^{19}$ whereas the velocity magnitude is around $\sim 1$), i.e.
\begin{equation}
    \mathbb{X}_{\psi, ij} \longleftarrow \frac{ \psi(\vec{x}_i; t_j)}{ \langle \psi(\vec{x}_i; 0)\rangle}
\end{equation}

The Randomized SVD, implemented in \textit{scikit-learn} \cite{scikit-learn}, is used to obtain a reduced representation in terms of the first $r$ principal components:
\begin{equation}
    \mathbb{U}_{\psi, r}\mathbbl{\Sigma}_{\psi, r}\mathbb{V}^*_{\psi,r} \approx \mathbb{X}_{\psi}
    \label{eqn: svd}
\end{equation}

The columns of the orthogonal matrix $\mathbb{U}_{\psi, r}\in\mathbb{R}^{\mathcal{N}_h\times r}$ represent the basis encoding the spatial information; the diagonal matrix $\mathbbl{\Sigma}_{\psi, r}\in\mathbb{R}^{r\times r}$ contains the singular values measuring the importance of each mode; the rows of the orthogonal matrix $\mathbb{V}_{\psi,r}\in\mathbb{R}^{N_t\times r}$ are the reduced coefficient encoding the dynamics of the snapshots. The rank $r$ is taken equal to 10 for all the 20 fields, adopting a truncation based on the singular values \cite{quarteroni2015reduced}, and the reduced dynamics at time $t_j$ are then collected into a reduced state space vector $\vec{v}_j\in\mathbb{R}^{r \cdot 20}$ being the temporal dynamics learnt by the SHRED architecture (Figure \ref{fig: shred}).

Focusing on the input of SHRED, the sensor locations can only be sampled in the reflector region (Figure \ref{fig: evol-geom}) and as the observable field the fast flux $\phi_1(\vec{x}; t)$ has been selected, given $\vec{x}\in \Omega$ and time $t\in[0,25]$ s, which is a sufficiently simple quantity to be measured in real scenarios. This particular choice is not a limitation of the method, as any of the fields can be selected as observed quantity (even though, in the real application, not all fields of interest could be easily measurable). The measurements have been synthetically generated from the OpenFOAM data computed with the model in Section \ref{sec: msfr}, exploiting the pyforce package developed by the authors \cite{riva2024multiphysics, ICAPP_plus2023} (\url{https://github.com/ERMETE-Lab/ROSE-pyforce}); in particular, the sensors are modelled as linear functionals $l_m(\cdot)=l(\cdot;\,\vec{x}_m, s)$ with a Gaussian kernel, as in \cite{maday_generalized_2015, ICAPP_plus2023}, each characterised by its the centre of mass $\vec{x}_m\in\Omega_{\text{refl}}$ and point spread $s$
\begin{equation}
    l_m(\phi_1(\vec{x});\,\vec{x}_m,\,s)=\int_\Omega \phi_1(\vec{x})\cdot K\cdot e^{\frac{-\|\vec{x}-\vec{x}_m\|_2^2}{2s^2}}\,d\Omega
    \label{eqn: sens-def}
\end{equation}
with $K$ defined such that $l_m(1; \,\vec{x}_m,\,s) = 1$ \cite{maday_generalized_2015}. For this application, the point spread has been taken equal to 0.025 as done in \cite{ICAPP_plus2023}. 

As already mentioned, the only available positions for the SHRED model are located in the reflector region $\Omega_{\text{refl}}$: this choice is intentional, as the aim of this work consists of developing a state estimation routine for the control and the monitoring of nuclear reactor and in-core measurements are quite hard to make. Additionally, this work also has the second objective of giving a measure of the uncertainty of the prediction, as the SHRED architecture is quick to train \cite{williams2022data, ebers2023leveraging} and it does not require powerful GPUs for the training phase and even personal computers can be used to perform quickly and reliably this task. This implies that several SHRED models can be trained with different random selection of sensors, such that different outputs, i.e. different reduced state space vectors $\vec{v}$, are produced, from which the sample mean and the associated standard deviation can be obtained. In this way, the predictions become more robust against random noise and this framework can be generalised to identify possible malfunctions in the reactors; if more than one SHRED model sees an upcoming accidental scenario, proper countermeasures can be undertaken based on more statistically consistent data.

\begin{figure}[t]
    \centering
    \includegraphics[width=1\linewidth]{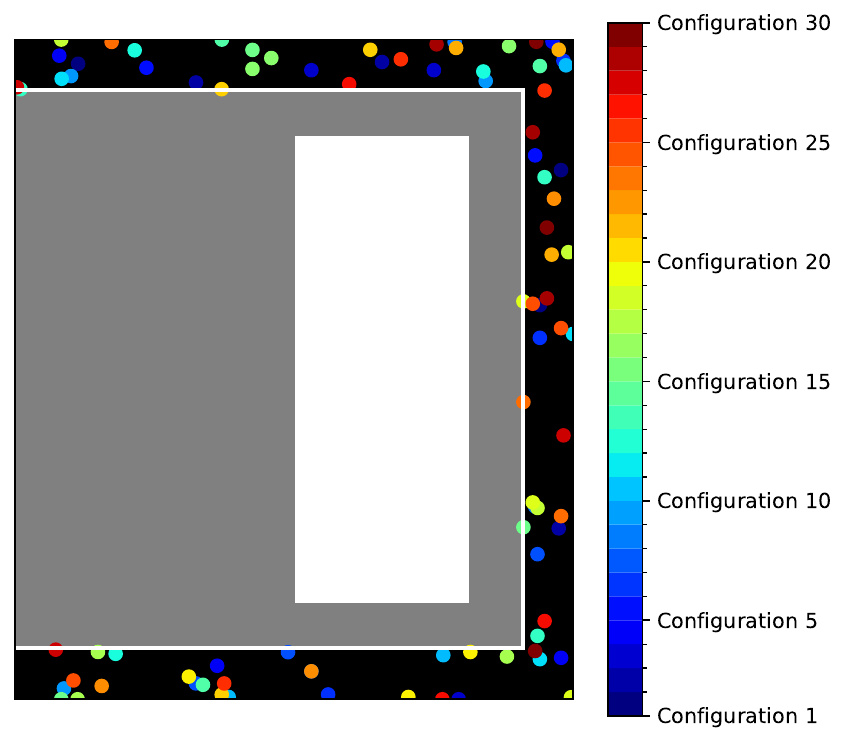}
    \caption{Random sensor configurations adopted in this work, placing them in the reflector region only (black zone); the gray area is the actual core. As expected, there is no discernible pattern in the positioning of the sensors in neither direction.}
    \label{fig: sensors}
\end{figure}

Therefore, in this work $L = 30$ SHRED models are trained using different sensor configurations: Figure \ref{fig: sensors} reports all the different positions for each configuration in the reflector region. The sensors have been randomly placed to show the reliability of the SHRED architecture and its capability to provide an accurate state estimation of the whole state space, even when measurements are randomly picked in the domain.

The capability of the SHRED architecture to retrieve the full state information from randomly selected sensors is of particular interest for the Nuclear Engineering community, where the problem of finding the optimal sensor placement is still an open issue \cite{argaud_sensor_2018, cannarile_novel_2018}. The proposed framework for Uncertainty Quantification based on the SHRED architecture can be easily integrated with algorithms for optimal sensor placement in the following way: the appropriate algorithm select the most significant locations; then, the SHRED is used to generate robust and reliable predictions by randomly choosing among the candidate locations. This will increase the monitoring capabilities of nuclear reactors, ensuring a safer system. Indeed, in the nuclear power plants industry, the selection of the location of sensors cannot be fully arbitrary, as regulator authorities require to have sensors placed in specific locations to monitor the most sensitive points of the engineering system (for example, where the temperature is expected to reach its maximum). Thus, the coupling between the SHRED architecture and an optimal sensor positioning algorithm will, on the one hand, make use of the potentiality of the SHRED architecture. In contrast, the positioning algorithm will ensure that locations are selected among those of interest from the regulatory and safety point of view.

Once the positions of the sensors have been chosen, the measures can be generated, i.e. the fast flux measurements $\vec{y}_{\phi_1}^k(t_j)\in\mathbb{R}^3$ of SHRED configuration $k$. To mimic the real scenario, these have been polluted with random zero-mean Gaussian noise $\epsilon\sim\mathcal{N}(0, \sigma^2)$
\begin{equation}
    \label{eq:noisymeas}
    y_{\phi_1, m}^k(t_j) = (1+ \epsilon)\cdot l_m^k(\phi_1(\vec{x}, t_j);\,\vec{x}_m,\,s) \qquad m=\{1, 2,3\}
\end{equation}
in which the standard deviation of the noise is considered to be 0.05, which is a reasonable value for the out-core of a MSFR \cite{brovchenko2013optimization}. The data have been split into three subsets (train, validation and test) for training and testing the neural network; then, the $k-$th SHRED model is trained in order to learn the input-output map between the measurements $\vec{y}_{\phi_1}^k(t)$ and the reduced state space vector $\vec{v}$. 

\section{Numerical Results}\label{sec: num-res}

\subsection{Learning The Dynamics}

\begin{figure*}[t]
    \centering
    \includegraphics[width=1\linewidth]{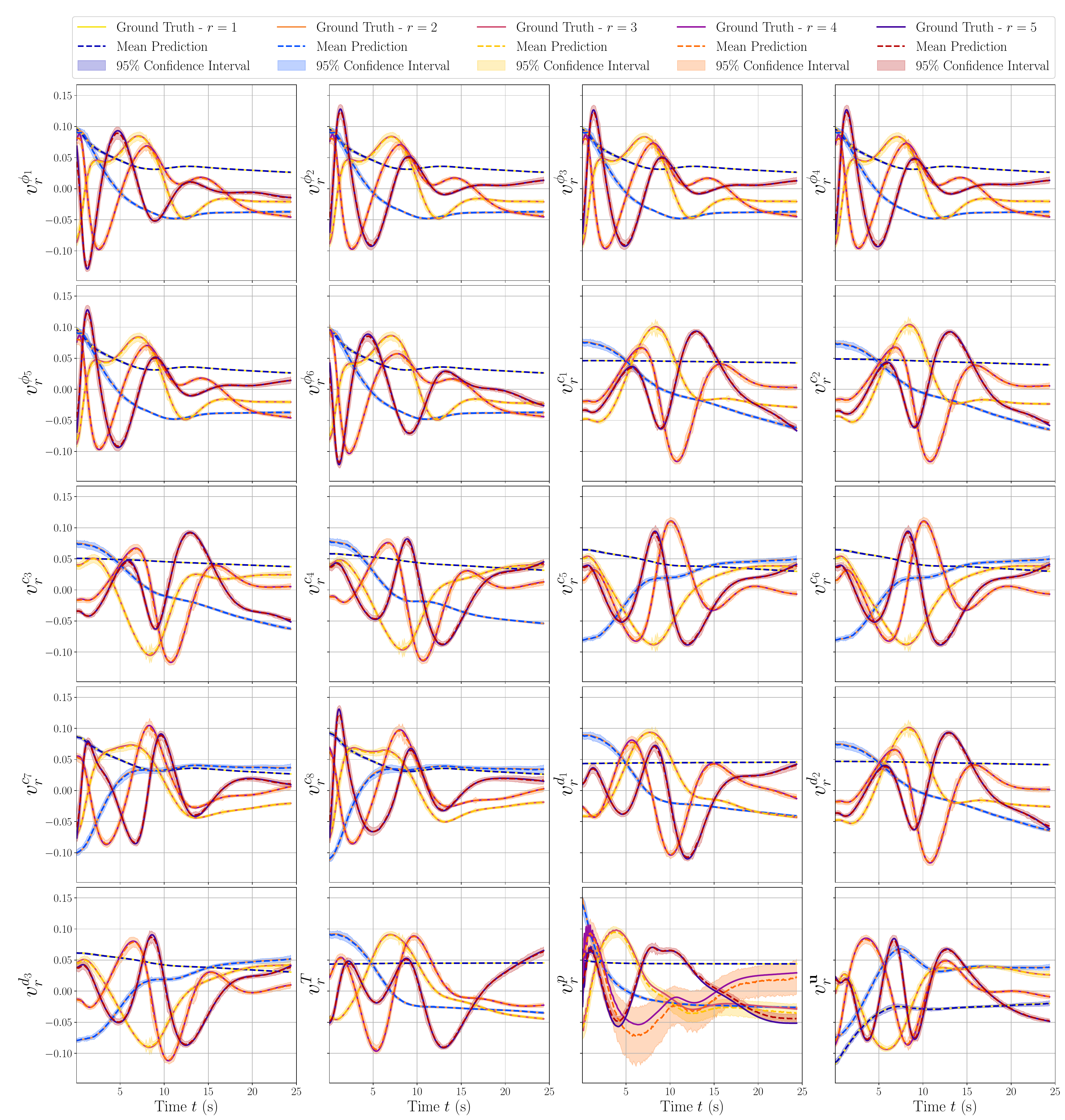}
    \caption{Comparison of the SHRED reconstruction of the reduced state space vector $\vec{v}$ with respect to the test dataset for the first 5 SVD coefficients of each field: continuous curves represent the mean of the SHRED models, the dashed lines are the ground truth (descending from the full-order data) and the shaded areas highlight the uncertainty regions for the SHRED models. The SHRED architecture is indeed able to correctly learn the dynamics underlying the considered accidental scenario, and by employing different SHRED models a mean prediction, statistically close to the actual value, can be achieved.}
    \label{fig: shred_dynamics_uq}
\end{figure*}

\begin{figure*}[t]
    \centering
    \includegraphics[width=0.9\linewidth]{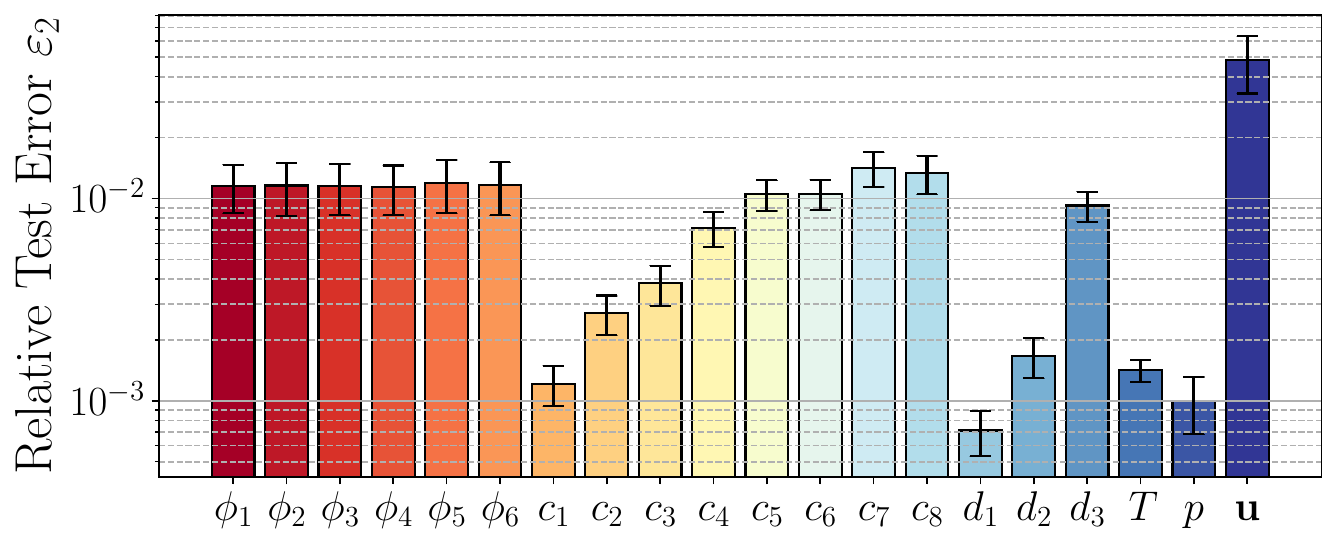}
    \caption{Average (with respect to the different configurations) relative reconstruction errors $\varepsilon_2 = \langle \varepsilon_2^k \rangle$ between the SHRED reconstruction and the FOM for the test set, along with the standard deviation pertaining to each variable of interest, measured in the $l_2$-norm. The velocity (vector) field proves to be the hardest one to reconstruct; still, the information coming from a single field ($\phi_1$) is sufficient to reconstruct the other fields with good accuracy.}
    \label{fig: rel-err-fom}
\end{figure*}

Since different SHRED models have been trained, different output predictions are retrieved, which can be used to obtain an estimate in terms of mean value and the standard deviation of each reduced coefficient $v_r^\psi$. Figure \ref{fig: shred_dynamics_uq} shows the first 5 SVD coefficients of each field of the state space, comparing the Ground Truth (solid lines), i.e. the Full Order Model (FOM), the SHRED mean value (dashed lines) and the associated confidence interval at 95\%. The SHRED output is in almost perfect agreement with the Ground Truth, highlighting that the LSTM and the Shallow Decoder are robust and reliable tools able to learn correctly the dynamics underlying this accidental scenario; the pressure coefficients are the ones characterised by the highest uncertainty; nevertheless, the employment of different SHRED models allows for a mean prediction which is statistically close to the actual value. The errors of each SHRED simulation oscillates slightly compared to the average value, due to the fact that sensors may be placed in points very close to the wall in which the temporal dynamics are less pronounced; however, the ensemble allows for a more robust prediction compared to the single simulation, making the output almost independent on the sensor placement. In terms of computational costs, each SHRED model takes about 1 minute of wall-clock time for the training phase on a workstation, and to get a new output, the associated computational cost is barely null these values for the training of SHRED models refers to a workstation with an Intel Core i7-9800X CPU with clock speed 3.80 GHz).

Now that the dynamics have been learnt, it is important to assess how the actual fields are predicted at the full-order level. Within the test set, the snapshots can be reconstructed using Eq. \eqref{eqn: svd}, assessing the reliability of the SHRED architecture coupled with the SVD. Let $\hat{\mathbb{X}}_{\psi, j}^k\in\mathbb{R}^{\mathcal{N}_h}$ be the reconstructed $j-$th snapshot for SHRED configuration $k$ of the generic field $\psi$ and let $\mathbb{X}_{\psi, j}\in\mathbb{R}^{\mathcal{N}_h}$ be the full-order solution; then, the relative error for the generic field $\psi$ on the test set at time $t_j$ for configuration $k$ can be defined as
\begin{equation}
    \varepsilon_{2,j}^{\psi, k} = \frac{\norma{\hat{\mathbb{X}}_{\psi, j}^k - \mathbb{X}_{\psi, j}}_2}{\norma{\mathbb{X}_{\psi, j}}_2}
\end{equation}
from which the average (in time) relative test error $\varepsilon_2^{\psi, k}$ of field $\psi$ for SHRED configuration $k$ can be computed as
\begin{equation}
    \varepsilon_{2}^{\psi, k} = \frac{1}{N_{t, test}} \sum_{j\in N_{t, test}}  \varepsilon_{2,j}^{\psi, k}
\end{equation}
from which the mean and the standard deviation considering all SHRED configurations can be computed.

Figure \ref{fig: rel-err-fom} shows a bar plot of the average relative test error $\varepsilon_{2}^{\psi}$ and the associated standard deviation for each field, measured in the $l_2-$norm. It can be observed that the information coming from the measurements of the fast flux $\phi_1$ is sufficient to obtain the same error level on the other group fluxes, and the error itself is kept around the measurement noise level (5\%); whereas the delayed groups (both precursors $c_k$ and decay heat $d_i$) are reconstructed with a low relative energy norm, as well as the temperature and the pressure: this means that the coupling between these fields and the observed one (the fast flux) is strong enough to provide a good reconstruction with sufficiently high accuracy. In the end, the most complex field to be reconstructed is the velocity field $\vec{u}$, which, in this case, is the one showing the slowest decay of the singular values of the SVD; likely, this field requires more SVD modes compared to the other ones to catch all the relevant information, including that of the smaller scales. In this case, for the sake of simplicity, the rank $r$ has been taken the same for all the different fields; this assumption can, however, be easily relaxed, allowing for different ranks $r$ according to the singular values decay of the particular field of interest. It is worth mentioning how this limit is not related to the SHRED architecture itself, but rather to the preliminary data compression step. 

I \subsection{Sensitivity Assessment on Sensors}

\begin{figure}
    \centering
    \includegraphics[width=1\linewidth]{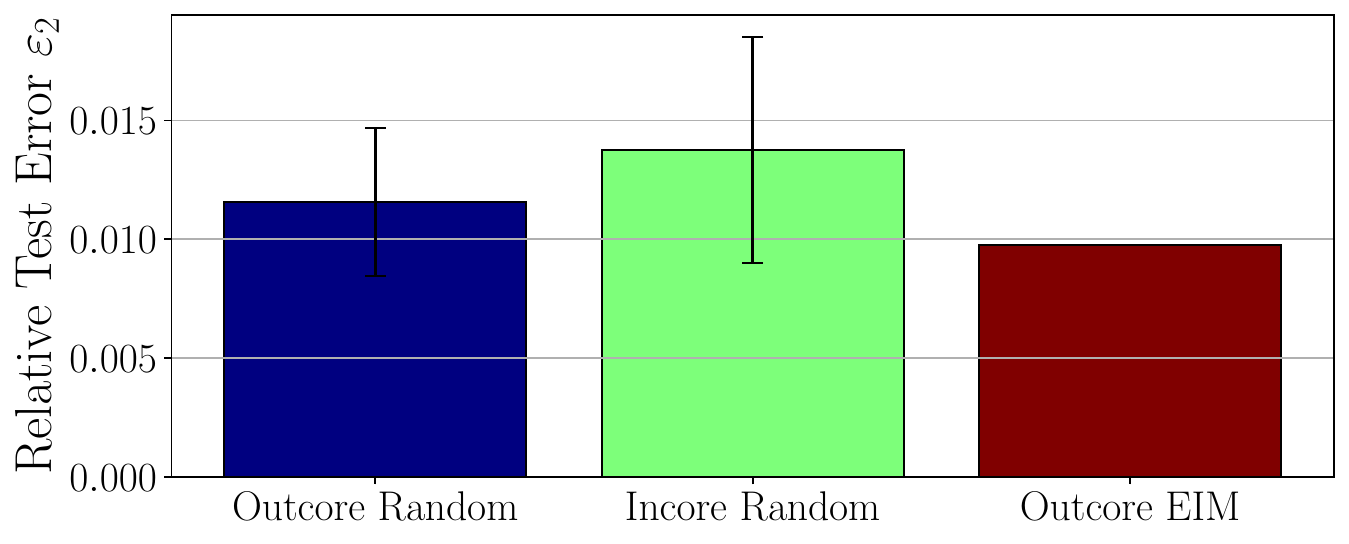}
    \caption{Comparison of the relative errors of the measured field $\phi_1$ using SHRED, with 3 different sensor configurations: randomly selected outcore (blue) and incore (green) and the first 3 sensor given by EIM (red) \cite{maday_eim_2008}. Results are very similar, indicating how SHRED is agnostic to sensor positioning and hierarchy.}
    \label{fig: comparison-3-sens}
\end{figure}

\begin{figure}
    \centering
    \includegraphics[width=1\linewidth]{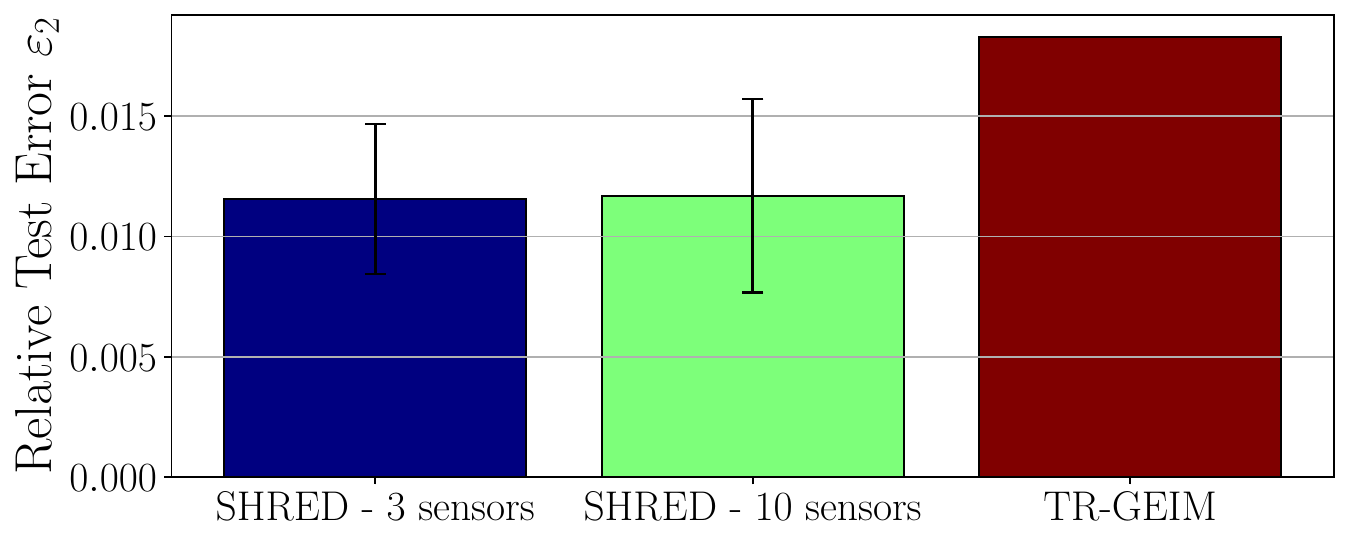}
    \caption{Comparison of the relative errors of the measured field $\phi_1$ using SHRED, with 3 (blue) and 10 (green) random sensors and the TR-GEIM  \cite{introini_stabilization_2023} algorithm (red), indicating how SHRED is independent of the number of available sensors, robust against noise, and comparable in performance with a state-of-the-art technique for sensor positioning in nuclear reactors.}
    \label{fig: comparison-10-sens}
\end{figure}

\textcolor{black}{Furthermore, a sensitivity analysis of the sensor placement constraints and the number of sensors has been performed. From a nuclear regulatory standpoint, placing sensor randomly could be hard to accept: therefore, it is worth comparing the overall output of SHRED with other reduced-order techniques that instead select sensors hierarchically, such as the Empirical Interpolation Method (EIM) \cite{maday_eim_2008}  its generalised version (GEIM) \cite{maday_generalized_2015}. Figure \ref{fig: comparison-3-sens} compares the performance of the SHRED architecture in terms of the average relative error for the measured field $\phi_1$ considering 3 sensors either randomly placed in the out-core region, in the core itself or selected as the first 3 sensors (in order of importance) given by the Empirical Interpolation Method (EIM) \cite{maday_eim_2008}: these configurations provide very similar results, and little improvements are coming from EIM sensors, highlight how the SHRED architecture is agnostic to sensor positioning and almost agnostic to sensor hierarchy \cite{williams2022data, kutz_shallow_2024}. Having as input the measures collected at the EIM locations increases the importance of the location itself and this may explain why the EIM error is a bit lower.} Conversely, the incore SHRED shows a slightly higher error than the outcore configuration, due to the noise level: higher noise levels are expected within the core, both due to the higher fluxes but also due to the more complex instrumentation electronics. In particular, referring to Equation \ref{eq:noisymeas}, the noise depends on the measurement value: as the fluxes are higher within the core, also the noise will be higher, leading to a slightly worse performance of the incore SHRED compared to the outcore one.

\textcolor{black}{The SHRED architecture with 3 randomly placed sensors in the reflector region has also been compared in Figure \ref{fig: comparison-10-sens} to the case in which 10 measures are used as input for the neural network and to the case in which the Tikhonov Regularised GEIM (TR-GEIM) \cite{introini_stabilization_2023}, a state-of-the-art non-intrusive state estimation technique adopted in nuclear reactors, has been used for selecting the sensor positions. This figure shows that SHRED is independent of the number of measurements used, highlighting the fact that the dynamics is correctly embedded in 3 sensors; moreover, the comparison with TR-GEIM underlines its robustness with respect to noise, being SHRED able to reconstruct not only the measured field, but also the un-observable quantities.}  \textcolor{black}{Additionally, as TR-GEIM selects sensor locations based on an hierarchical algorithm, the fact that SHRED shows comparable, and even better, performances confirms how the SHRED architecture can go beyond the logic of optimising the sensor positioning, thus making it possible to select the best possible location based on economic and accessibility considerations.}

\subsection{SHRED Reconstruction}

\begin{figure}[t]
    \centering
    \includegraphics[width=1\linewidth]{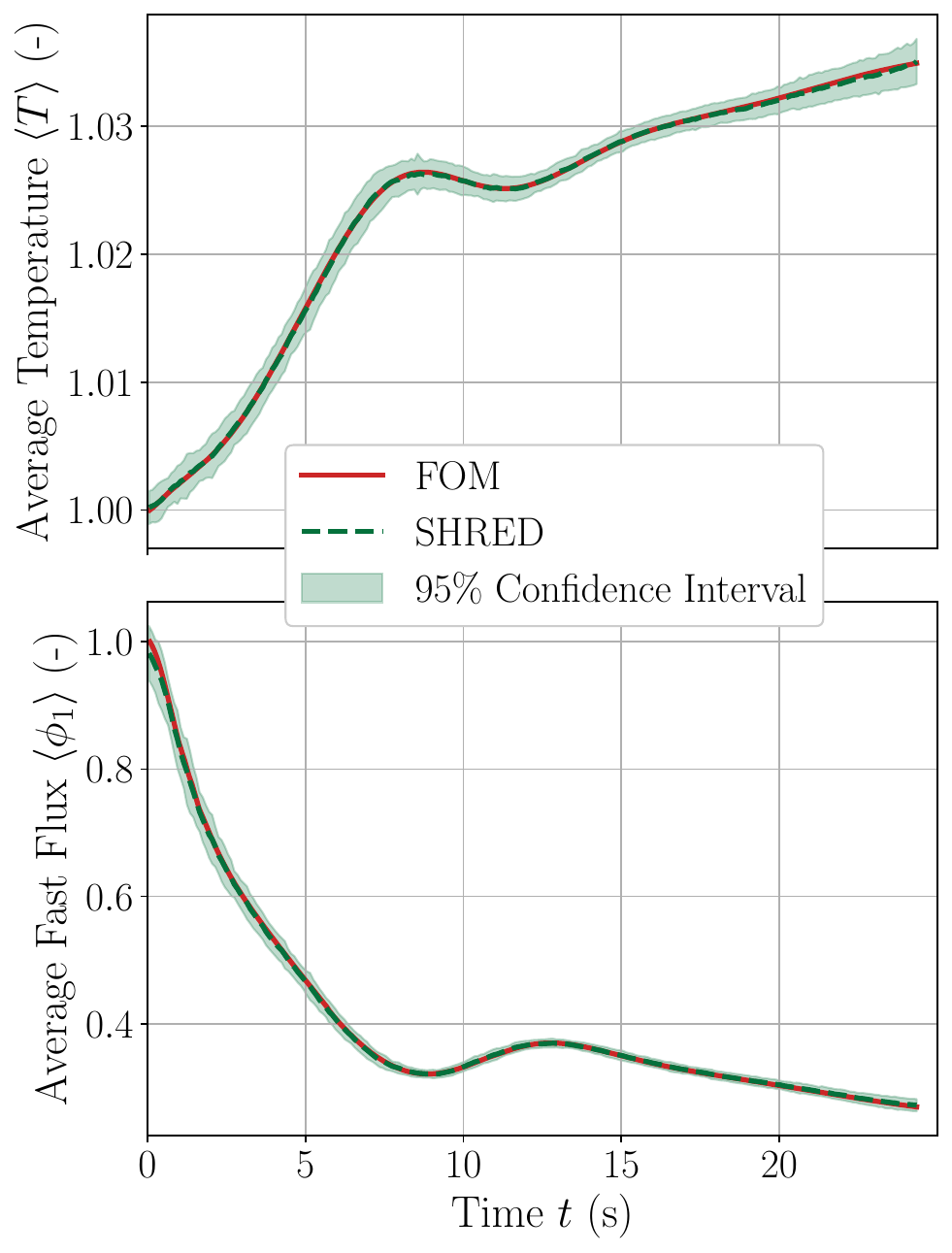}
    \caption{Comparison of the full-order and the SHRED prediction for the average temperature of whole domain (including reflector) and average fast flux, normalised to the initial value and with confidence interval 95\%, showing almost perfect agreement.}
    \label{fig: shred-global-qties}
\end{figure}

\begin{figure*}[t]
    \centering
        \begin{overpic}[width=1\linewidth]{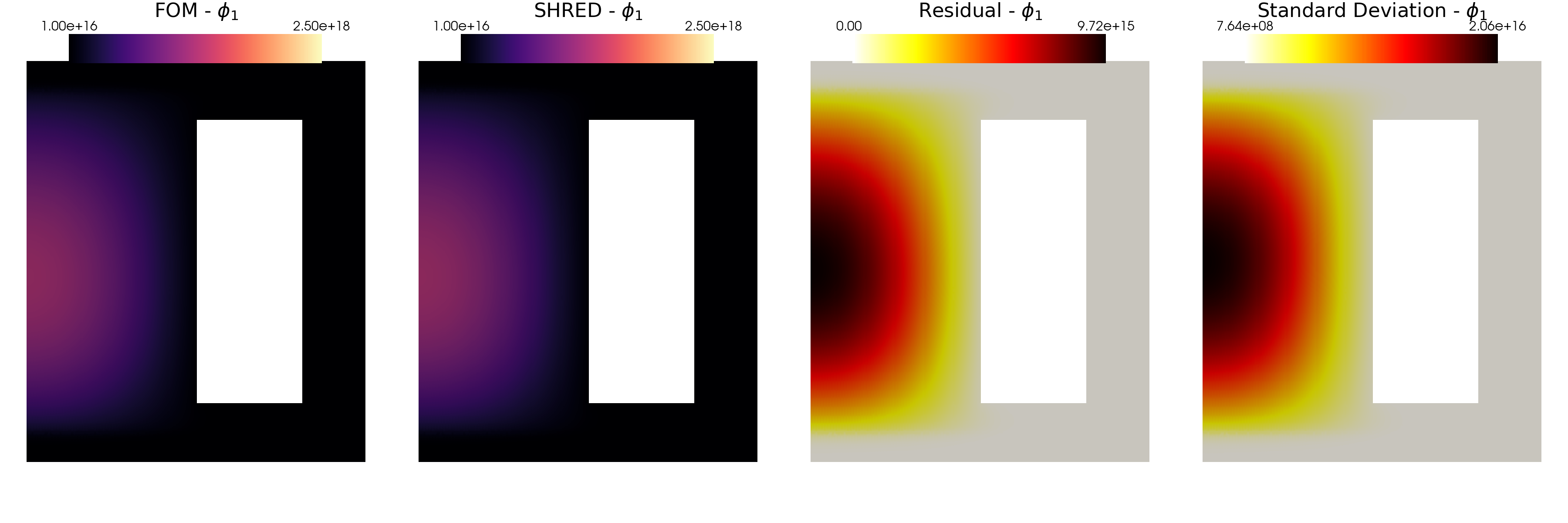}
        \put(0,32){(a)}   
        \end{overpic}
        \begin{overpic}[width=1\linewidth]{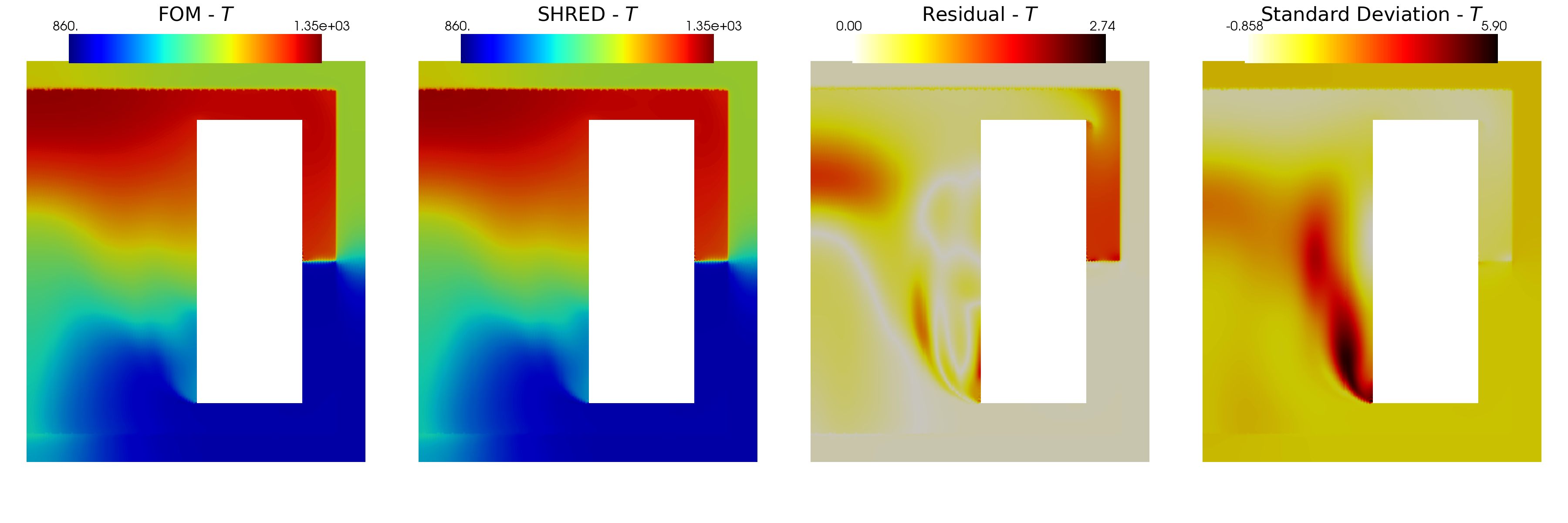}
        \put(0,32){(b)} 
        \end{overpic}
        \begin{overpic}[width=1\linewidth]{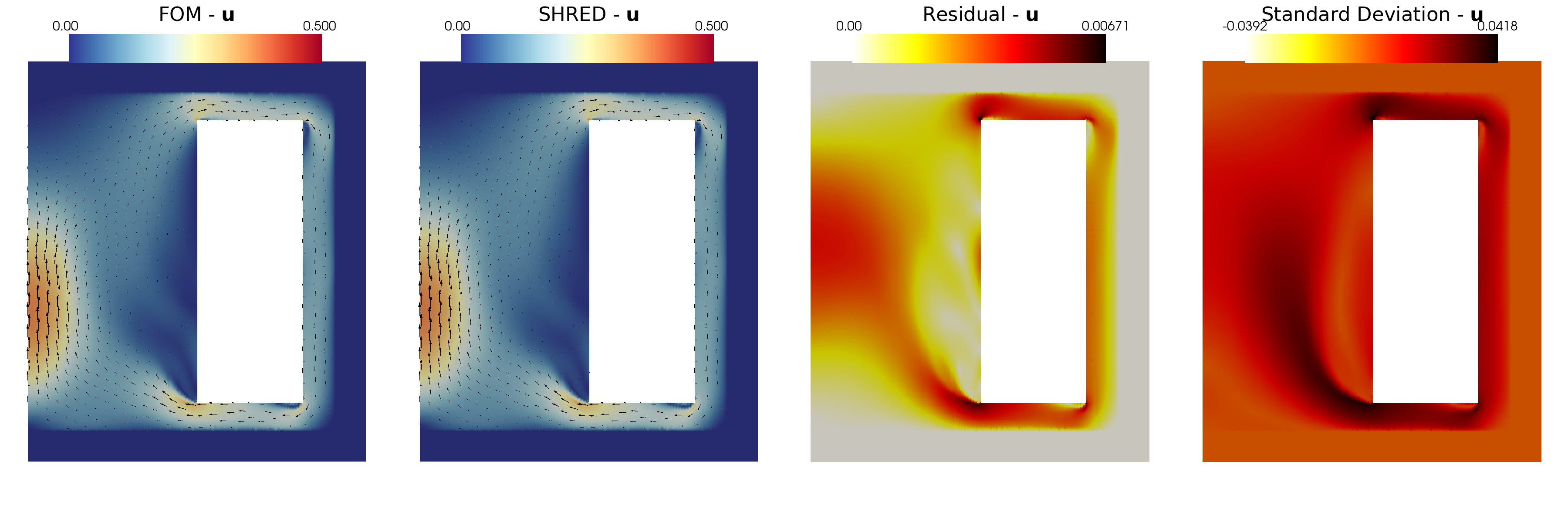}
        \put(0,32){(c)} 
        \put(84,1.){Time $t=24.35$ s.}
        \end{overpic}
    \caption{Contour plots near the end of the transient (for the last time instant in the test set) of the observed field $\phi_1$ (a), the temperature $T$ (b) and the velocity $\vec{u}$ (c). From left to right, there is the full-order solution, the mean of the different SHRED models, the associated residual field and the standard deviation of the different SHRED models. The SHRED model provides a correct state estimation both of the observable and the un-observable ones; the right-most column, showing the standard deviation, allows to see the dominant structures cut off by the SVD, highlighting where the estimation is poorer.}
    \label{fig: contours}
\end{figure*}

This framework can be very useful to obtain estimates of integral quantities typically used to monitor and control the nuclear reactor, such as the average (or maximum) temperature $\langle T \rangle$ and the core power (directly related to the average neutron flux). Figure \ref{fig: shred-global-qties} reports the average temperature and the average fast flux, normalised to the initial value of the FOM and the mean SHRED reconstruction and the related uncertainty given by the different sensor configurations. The agreement is almost perfect for both quantities, showing that the SHRED is very accurate in estimating integral quantities describing on average what is occurring inside the nuclear reactor.

In the end, Figure \ref{fig: contours} shows some contour plots of the SHRED prediction at the last time step, compared against the FOM and along with the associated residual field $r_\psi$, defined as the absolute difference between the FOM ($\psi$) and the SHRED prediction ($\hat{\psi}$), i.e.
\begin{equation}
    r_{\psi}(\vec{x};t) = \left| \psi (\vec{x};t)- \hat{\psi}(\vec{x};t) \right|
\end{equation}
\textcolor{black}{In addition, the standard deviation, with respect to the SHRED outputs, of the reconstructed fields is show, highlighting the zones with higher variation and thus the dominant structures cut off by the SVD.}
The SHRED model can provide a correct state estimation of the observable field $\phi_1$ and the un-observable ones, such as temperature $T$ and velocity $\vec{u}$. The latter is the most difficult one to reconstruct, nevertheless, the residual field in terms of magnitude is reasonably low. Some videos of the whole transient can be found at this \href{https://www.youtube.com/playlist?list=PLSH6dleR9E1C1u2qHTxDDqLlc-q5iN_Z7}{link}.

\section{Conclusions}\label{sec: conclusions}

This work demonstrates the application of the SHallow REcurrent Decoder network to state estimation in advanced nuclear reactors, considering the test case of the Generation-IV Molten Salt Fast Reactor with only access to three out-core sensor measurements of the fast flux. In particular, the EVOL geometry was selected as the case study, considering the Unprotected Loss of Fuel Flow accidental scenario in which the external pumps lose power exponentially. The SHRED architecture is used to provide an estimation of the entire state space vector, which includes the energy fluxes, the neutron precursors, the decay heat groups and the thermal-hydraulics fields for a total of 20 coupled variables, using as input the out-core sensor measurements of the fast flux only, that is, collected with sensors placed in the external solid reflector region to mimic the actual reactor in which it will be impossible to have in-core sensors due to the liquid nature of the fuel.

In particular, several SHRED models have been trained considering different sensor configurations (in the reflector only), and hence different input measures: in this way, in addition to the full state estimation, a quantification of the associated uncertainty can be obtained, making the prediction more robust and reliable against noisy measurements. The results presented in this work are promising and show how the SHRED architecture can be used to reconstruct the state of a nuclear reactor starting from partial out-core observations.

As a first application of the SHRED architecture to nuclear cases, this paper considers a single accidental transient for the reconstruction; in the future, this same methodology will be used to identify the different scenarios in a classification paradigm, adding another fundamental algorithm for the development of digital twins of nuclear reactors. Moreover, this technique will be extended to possibly correct/update mathematical models from the local observation or to combine different levels of fidelity to improve the accuracy, both in the design and the operation phase.

\section*{Code}  

The code and data (compressed) are available at: \url{https://github.com/ERMETE-Lab/NuSHRED}.

\section*{Acknowledgments} 
The contribution of Nathan Kutz was supported in part by the US National Science Foundation (NSF) AI Institute for Dynamical Systems (dynamicsai.org), grant 2112085.

\bibliographystyle{unsrt}
\bibliography{bibliography}

\begin{thebibliography}{10}

\bibitem{GenIV-RoadMap}
{Generation IV International Forum}.
\newblock {Technology Roadmap Update for Generation IV Nuclear Energy Systems}, 2014.

\bibitem{brovchenko2013optimization}
Mariya Brovchenko, Elsa Merle~Lucotte, Herv{\'e} Rouch, Fabio Alcaro, M~Allibert, M~Aufiero, Antonio Cammi, S~Dulla, O~Feynberg, L~Frima, et~al.
\newblock Optimization of the pre-conceptual design of the msfr, 2013.

\bibitem{DuderstadtHamilton}
James~J. Duderstadt and Louis~J. Hamilton.
\newblock {\em Nuclear reactor analysis}.
\newblock Wiley New York, 1976.

\bibitem{international2023iaea}
International Atomic~Energy Agency.
\newblock Nuclear–renewable hybrid energy systems.
\newblock Technical Report NR-T-1.24, IAEA, Vienna, 2023.

\bibitem{RUTH2014684}
Mark~F. Ruth, Owen~R. Zinaman, Mark Antkowiak, Richard~D. Boardman, Robert~S. Cherry, and Morgan~D. Bazilian.
\newblock Nuclear-renewable hybrid energy systems: Opportunities, interconnections, and needs.
\newblock {\em Energy Conversion and Management}, 78:684--694, 2014.

\bibitem{Grieves_DT}
Michael~W. Grieves.
\newblock {\em Virtually Intelligent Product Systems: Digital and Physical Twins}, pages 175--200.
\newblock Progress in Astronautics and Aeronauting, 2019.

\bibitem{mohanty_physics-infused_2021}
Subhasish Mohanty and Richard Vilim.
\newblock Physics-{Infused} {AI}/{ML} {Based} {Digital}-{Twin} {Framework} for {Flow}-{Induced}-{Vibration} {Damage} {Prediction} in a {Nuclear} {Reactor} {Heat} {Exchanger}.
\newblock Technical Report ANL/NSE-21/8, 1830413, 172195, Argonne National Laboratory, September 2021.

\bibitem{mohanty_development_2021}
Subhasish Mohanty and Joseph Listwan.
\newblock Development of {Digital} {Twin} {Predictive} {Model} for {PWR} {Components}: {Updates} on {Multi} {Times} {Series} {Temperature} {Prediction} {Using} {Recurrent} {Neural} {Network}, {DMW} {Fatigue} {Tests}, {System} {Level} {Thermal}-{Mechanical}-{Stress} {Analysis}.
\newblock Technical Report ANL/LWRS-21/02, 1822853, 171255, Argonne National Laboratory, September 2021.

\bibitem{gong_data-enabled_2022}
Helin Gong, Sibo Cheng, Zhang Chen, and Qing Li.
\newblock Data-{Enabled} {Physics}-{Informed} {Machine} {Learning} for {Reduced}-{Order} {Modeling} {Digital} {Twin}: {Application} to {Nuclear} {Reactor} {Physics}.
\newblock {\em Nuclear Science and Engineering}, 196(6):668--693, June 2022.
\newblock Publisher: Taylor \& Francis \_eprint: https://doi.org/10.1080/00295639.2021.2014752.

\bibitem{argaud_sensor_2018}
J.-P. Argaud, B~Bouriquet, F~de~Caso, H~Gong, Y~Maday, and O~Mula.
\newblock Sensor placement in nuclear reactors based on the generalized empirical interpolation method.
\newblock {\em J. Comput. Phys.}, 363:354--370, 2018.

\bibitem{gong_parameter_2023}
Helin Gong, Tao Zhu, Zhang Chen, Yaping Wan, and Qing Li.
\newblock Parameter identification and state estimation for nuclear reactor operation digital twin.
\newblock {\em Annals of Nuclear Energy}, 180:109497, January 2023.

\bibitem{gong_efficient_2022}
Helin Gong, Sibo Cheng, Zhang Chen, Qing Li, César Quilodrán-Casas, Dunhui Xiao, and Rossella Arcucci.
\newblock An efficient digital twin based on machine learning {SVD} autoencoder and generalised latent assimilation for nuclear reactor physics.
\newblock {\em Annals of Nuclear Energy}, 179:109431, December 2022.

\bibitem{ICAPP_plus2023}
Antonio Cammi, Stefano Riva, Carolina Introini, Lorenzo Loi, and Enrico Padovani.
\newblock {Data-driven model order reduction for sensor positioning and indirect reconstruction with noisy data: Application to a Circulating Fuel Reactor}.
\newblock {\em Nuclear Engineering and Design}, 421:113105, 2024.

\bibitem{PHYSOR24_MSFR_outcore}
Stefano Riva, Sophie Deanesi, Carolina Introini, Stefano Lorenzi, and Antonio Cammi.
\newblock {Neutron Flux Reconstruction from Out-Core Sparse Measurements using Data-Driven Reduced Order Modelling}.
\newblock In {\em International Conference on Physics of Reactors (PHYSOR24)}, San Francisco, USA, April 2024.

\bibitem{Kapteyn_Willcox_2020}
Michael~G. Kapteyn, David~J. Knezevic, and Karen Willcox.
\newblock Toward predictive digital twins via component-based reduced-order models and interpretable machine learning.
\newblock In {\em AIAA Scitech 2020 Forum}, 2020.

\bibitem{lassila_model_2014}
Toni Lassila, Andrea Manzoni, Alfio Quarteroni, and Gianluigi Rozza.
\newblock Model {Order} {Reduction} in {Fluid} {Dynamics}: {Challenges} and {Perspectives}.
\newblock In {\em Reduced {Order} {Methods} for {Modeling} and {Computational} {Reduction}}, pages 235--273. Springer International Publishing, Cham, 2014.

\bibitem{rozza_model_2020}
Gianluigi Rozza, Martin Hess, Giovanni Stabile, Marco Tezzele, Francesco Ballarin, Carmen Gräßle, Michael Hinze, Stefan Volkwein, Francisco Chinesta, Pierre Ladeveze, Yvon Maday, Anthony Patera, and J~Farhat~Char.
\newblock {\em Model {Order} {Reduction}: {Volume} 2: {Snapshot}-{Based} {Methods} and {Algorithms}}.
\newblock De Gruyter, 2020.

\bibitem{quarteroni2015reduced}
A~Quarteroni, A~Manzoni, and F~Negri.
\newblock {\em {Reduced Basis Methods for Partial Differential Equations: An Introduction}}.
\newblock UNITEXT. Springer International Publishing, 2015.

\bibitem{brunton_data-driven_2022}
Steven~L Brunton and J~Nathan Kutz.
\newblock {\em Data-{Driven} {Science} and {Engineering}: {Machine} {Learning}, {Dynamical} {Systems}, and {Control}}.
\newblock Cambridge University Press, USA, 2nd edition, 2022.

\bibitem{maday_generalized_2015}
Y~Maday, O~Mula, A~T Patera, and M~Yano.
\newblock The {Generalized} {Empirical} {Interpolation} {Method}: {Stability} theory on {Hilbert} spaces with an application to the {Stokes} equation.
\newblock {\em Computer Methods in Applied Mechanics and Engineering}, 287:310--334, 2015.
\newblock Publisher: Elsevier B.V.

\bibitem{maday_parameterized-background_2014}
Yvon Maday, Anthony Patera, James Penn, and Masayuki Yano.
\newblock A parameterized-background data-weak approach to variational data assimilation: formulation, analysis, and application to acoustics.
\newblock {\em Int. J. Numer. Methods Eng.}, 102, 2014.

\bibitem{williams2022data}
Jan Williams, Olivia Zahn, and J~Nathan Kutz.
\newblock Data-driven sensor placement with shallow decoder networks.
\newblock {\em arXiv preprint arXiv:2202.05330}, 2022.

\bibitem{binev_greedy_2018}
Peter Binev, Albert Cohen, Olga Mula, and James Nichols.
\newblock Greedy {Algorithms} for {Optimal} {Measurements} {Selection} in {State} {Estimation} {Using} {Reduced} {Models}.
\newblock {\em SIAM/ASA Journal on Uncertainty Quantification}, 6(3):1101--1126, January 2018.
\newblock Publisher: Society for Industrial and Applied Mathematics.

\bibitem{Manohar2018}
Krithika Manohar, Bingni~W. Brunton, J.~Nathan Kutz, and Steven~L. Brunton.
\newblock Data-driven sparse sensor placement for reconstruction: Demonstrating the benefits of exploiting known patterns.
\newblock {\em IEEE Control Systems Magazine}, 38(3):63--86, 2018.

\bibitem{INTROINI2023109538}
Carolina Introini, Stefano Riva, Stefano Lorenzi, Simone Cavalleri, and Antonio Cammi.
\newblock Non-intrusive system state reconstruction from indirect measurements: A novel approach based on hybrid data assimilation methods.
\newblock {\em Annals of Nuclear Energy}, 182:109538, 2023.

\bibitem{ebers2023leveraging}
Megan~R Ebers, Jan~P Williams, Katherine~M Steele, and J~Nathan Kutz.
\newblock Leveraging arbitrary mobile sensor trajectories with shallow recurrent decoder networks for full-state reconstruction.
\newblock {\em arXiv preprint arXiv:2307.11793}, 2023.

\bibitem{kutz_shallow_2024}
J.~Nathan Kutz, Maryam Reza, Farbod Faraji, and Aaron Knoll.
\newblock Shallow {Recurrent} {Decoder} for {Reduced} {Order} {Modeling} of {Plasma} {Dynamics}, May 2024.
\newblock arXiv:2405.11955 [nlin, physics:physics].

\bibitem{aufiero2014development}
Manuele Aufiero.
\newblock {\em {Development of Advanced Simulation Tools for Circulating-Fuel Nuclear Reactors}}.
\newblock PhD thesis, Politecnico di Milano, 03 2014.

\bibitem{hochreiter1997long}
Sepp Hochreiter and J{\"u}rgen Schmidhuber.
\newblock Long short-term memory.
\newblock {\em Neural computation}, 9(8):1735--1780, 1997.

\bibitem{erichson2020shallow}
N~Benjamin Erichson, Lionel Mathelin, Zhewei Yao, Steven~L Brunton, Michael~W Mahoney, and J~Nathan Kutz.
\newblock Shallow neural networks for fluid flow reconstruction with limited sensors.
\newblock {\em Proceedings of the Royal Society A}, 476(2238):20200097, 2020.

\bibitem{lipton2015critical}
Zachary~C Lipton, John Berkowitz, and Charles Elkan.
\newblock A critical review of recurrent neural networks for sequence learning.
\newblock {\em arXiv preprint arXiv:1506.00019}, 2015.

\bibitem{yu2019review}
Yong Yu, Xiaosheng Si, Changhua Hu, and Jianxun Zhang.
\newblock A review of recurrent neural networks: Lstm cells and network architectures.
\newblock {\em Neural computation}, 31(7):1235--1270, 2019.

\bibitem{takens1981lnm}
F~Takens.
\newblock Detecting strange attractors in turbulence.
\newblock {\em Lecture Notes in Mathematics}, 898:366--381, 1981.

\bibitem{gong_reactor_2024}
He-Lin Gong, Han Li, Dunhui Xiao, and Sibo Cheng.
\newblock Reactor field reconstruction from sparse and movable sensors using {Voronoi} tessellation-assisted convolutional neural networks.
\newblock {\em Nuclear Science and Techniques}, 35(5):43, April 2024.

\bibitem{pytorch}
Jason Ansel, Edward Yang, Horace He, Natalia Gimelshein, Animesh Jain, Michael Voznesensky, Bin Bao, Peter Bell, David Berard, Evgeni Burovski, Geeta Chauhan, Anjali Chourdia, Will Constable, Alban Desmaison, Zachary DeVito, Elias Ellison, Will Feng, Jiong Gong, Michael Gschwind, Brian Hirsh, Sherlock Huang, Kshiteej Kalambarkar, Laurent Kirsch, Michael Lazos, Mario Lezcano, Yanbo Liang, Jason Liang, Yinghai Lu, C.~K. Luk, Bert Maher, Yunjie Pan, Christian Puhrsch, Matthias Reso, Mark Saroufim, Marcos~Yukio Siraichi, Helen Suk, Shunting Zhang, Michael Suo, Phil Tillet, Xu~Zhao, Eikan Wang, Keren Zhou, Richard Zou, Xiaodong Wang, Ajit Mathews, William Wen, Gregory Chanan, Peng Wu, and Soumith Chintala.
\newblock Pytorch 2: Faster machine learning through dynamic python bytecode transformation and graph compilation.
\newblock In {\em Proceedings of the 29th ACM International Conference on Architectural Support for Programming Languages and Operating Systems, Volume 2}, ASPLOS '24, page 929–947, New York, NY, USA, 2024. Association for Computing Machinery.

\bibitem{serp_molten_2014}
Jérôme Serp, Michel Allibert, Ondřej Beneš, Sylvie Delpech, Olga Feynberg, Véronique Ghetta, Daniel Heuer, David Holcomb, Victor Ignatiev, Jan~Leen Kloosterman, Lelio Luzzi, Elsa Merle-Lucotte, Jan Uhlíř, Ritsuo Yoshioka, and Dai Zhimin.
\newblock The molten salt reactor ({MSR}) in generation {IV}: {Overview} and perspectives.
\newblock {\em Progress in Nuclear Energy}, 77:308--319, November 2014.

\bibitem{OpenMC}
Paul~K. Romano, Nicholas~E. Horelik, Bryan~R. Herman, Adam~G. Nelson, Benoit Forget, and Kord Smith.
\newblock {OpenMC: A state-of-the-art Monte Carlo code for research and development}.
\newblock {\em Annals of Nuclear Energy}, 82:90--97, 2015.
\newblock Joint International Conference on Supercomputing in Nuclear Applications and Monte Carlo 2013, SNA + MC 2013. Pluri- and Trans-disciplinarity, Towards New Modeling and Numerical Simulation Paradigms.

\bibitem{scikit-learn}
F.~Pedregosa, G.~Varoquaux, A.~Gramfort, V.~Michel, B.~Thirion, O.~Grisel, M.~Blondel, P.~Prettenhofer, R.~Weiss, V.~Dubourg, J.~Vanderplas, A.~Passos, D.~Cournapeau, M.~Brucher, M.~Perrot, and E.~Duchesnay.
\newblock Scikit-learn: Machine learning in {P}ython.
\newblock {\em Journal of Machine Learning Research}, 12:2825--2830, 2011.

\bibitem{riva2024multiphysics}
Stefano Riva, Carolina Introini, and Antonio Cammi.
\newblock Multi-physics model bias correction with data-driven reduced order techniques: Application to nuclear case studies.
\newblock {\em Applied Mathematical Modelling}, 135:243--268, 2024.

\bibitem{cannarile_novel_2018}
Francesco Cannarile, Piero Baraldi, Pierluigi Colombo, and Enrico Zio.
\newblock A {Novel} {Method} for {Sensor} {Data} {Validation} based on the analysis of {Wavelet} {Transform} {Scalograms}.
\newblock {\em International Journal of Prognostics and Health Management}, 9(1), 2018.
\newblock Number: 1.

\bibitem{maday_eim_2008}
Yvon Maday, Ngoc Nguyen, Anthony Patera, and George Shu~Heng Pau.
\newblock A general multipurpose interpolation procedure: {The} magic points.
\newblock {\em Communications on Pure and Applied Analysis}, 8, 2008.

\bibitem{introini_stabilization_2023}
Carolina Introini, Simone Cavalleri, Stefano Lorenzi, Stefano Riva, and Antonio Cammi.
\newblock Stabilization of {Generalized} {Empirical} {Interpolation} {Method} ({GEIM}) in presence of noise: {A} novel approach based on {Tikhonov} regularization.
\newblock {\em Computer Methods in Applied Mechanics and Engineering}, 404:115773, 2023.

\bibitem{versteeg2007introduction}
H~K Versteeg and W~Malalasekera.
\newblock {\em {An Introduction to Computational Fluid Dynamics: The Finite Volume Method}}.
\newblock Pearson Education Limited, 2007.

\end{thebibliography}

\clearpage

\section*{List of Symbols}
\nomenclature[L]{$\vec{y}$}{Measurement vector}
\nomenclature[L]{$t$}{Time}
\nomenclature[L]{$\Delta t$}{Time Step}
\nomenclature[L]{$N_t$}{Number of time snapshots}
\nomenclature[L]{$v_n$}{Reduced/Modal coefficient of rank $n$}
\nomenclature[L]{$r$}{Rank of the SVD}
\nomenclature[L]{$\dot{Q}$}{Pump Flow Rate}
\nomenclature[L]{$P$}{Core Power}
\nomenclature[L]{$T$}{Temperature}
\nomenclature[L]{$p$}{Pressure}
\nomenclature[L]{$\vec{u}$}{Velocity vector}
\nomenclature[L]{$\mathcal{V}$}{Full-Order state space}
\nomenclature[L]{$\mathcal{N}_h$}{Spatial degrees of freedom}
\nomenclature[L]{$\mathbb{X}_\psi$}{Snapshot matrix for generic field $\psi$}
\nomenclature[L]{$\hat{\mathbb{X}}_\psi$}{Reconstructed Snapshot matrix with SHRED for generic field $\psi$}
\nomenclature[L]{$\mathbb{U}_\psi$}{SVD basis for generic field $\psi$}
\nomenclature[L]{$\mathbbl{\Sigma}_\psi$}{SVD singular values for generic field $\psi$}
\nomenclature[L]{$\mathbb{V}_\psi$}{SVD reduced dynamics for generic field $\psi$}
\nomenclature[L]{$\vec{v}_j$}{Reduced state space vector at time $t_j$}
\nomenclature[L]{$\vec{x}$}{Space coordinate}
\nomenclature[L]{$l_m$}{$m-$th functional representing the sensor}
\nomenclature[L]{$s$}{Point-spread of the sensor}
\nomenclature[L]{$K$}{Normalisation constant for the sensor}
\nomenclature[L]{$L$}{Number of SHRED models}
\nomenclature[L]{$r_\psi$}{Residual field for generic field $\psi$}

\nomenclature[G]{$\Omega$}{Physical Domain}
\nomenclature[G]{$\kappa-\varepsilon$}{Turbulent Kinetic Energy and Turbulent Dissipation Rate}
\nomenclature[G]{$\phi_g$}{$g$-th Neutron group Flux}
\nomenclature[G]{$c_k$}{$k$-th precursors group}
\nomenclature[G]{$d_i$}{$i$-th decay heat group}
\nomenclature[G]{$\psi$}{Generic Field}
\nomenclature[G]{$\hat{\psi}$}{SHRED reconstruction of a Generic Field}
\nomenclature[G]{$\epsilon$}{Random Noise}
\nomenclature[G]{$\varepsilon_2$}{Relative Error between FOM and SHRED in energy norm}

\nomenclature[A]{DT}{Digital Twin}
\nomenclature[A]{MSFR}{Molten Salt Fast Reactor}
\nomenclature[A]{ROM}{Reduced Order Modelling}
\nomenclature[A]{PDE}{Partial Differential Equation}
\nomenclature[A]{POD}{Proper Orthogonal Decomposition}
\nomenclature[A]{SHRED}{SHallow REcurrent Decoder}
\nomenclature[A]{EVOL}{Evaluation and Viability of Liquid Fuel Fast Reactor System}
\nomenclature[A]{LSTM}{Long Short-Term Memory}
\nomenclature[A]{SDN}{Shallow Decoder Network}
\nomenclature[A]{SVD}{Singular Value Decomposition}
\nomenclature[A]{RANS}{Reynolds-Averaged Navier-Stokes}
\nomenclature[A]{ULOFF}{Unprotected Loss of Fuel Flow}
\nomenclature[A]{FOM}{Full Order Model}

\nomenclature[L]{$\vec{g}$}{Gravity}
\nomenclature[L]{$c_p$}{Specific Heat Capacity at constant pressure}
\nomenclature[L]{$q'''$}{Power Density}
\nomenclature[L]{$\bar{E}_f$}{Average Energy released by Fission}
\nomenclature[L]{$D_{n,g}$}{Neutron Diffusion coefficient of group $g$}
\nomenclature[L]{$S_{g}$}{Neutron Source for group $g$}
\nomenclature[L]{$S_{p,g}$}{Prompt Neutron Source from Fission in group $g$}
\nomenclature[L]{$S_{d}$}{Delayed Neutron Source}
\nomenclature[L]{$S_{s,g}$}{Source from Scattering Neutrons}
\nomenclature[L]{$S_{g}$}{Neutron Source for group $g$}
\nomenclature[L]{$A_{r,g}^0$}{Proportionality Constant for Doppler Broadening for group $g$}

\nomenclature[G]{$\nu_{\text{eff}}$}{Effective kinematic viscosity}
\nomenclature[G]{$\alpha_{\text{eff}}$}{Effective thermal diffusivity}
\nomenclature[G]{$\rho$}{Density}
\nomenclature[G]{$\beta_T$}{Thermal Expansion Coefficient}
\nomenclature[G]{$\beta_h$}{Delay Heat Fraction}
\nomenclature[G]{$\beta$}{Delayed Neutron Fraction}
\nomenclature[G]{$\chi_{p,g}$}{Prompt Neutron Spectrum of group $g$}
\nomenclature[G]{$\chi_{d,g}$}{Delayed Neutron Spectrum of group $g$}
\nomenclature[G]{$\Sigma_{f,g}$}{Fission Cross Section of group $g$}
\nomenclature[G]{$\Sigma_{r,g}$}{Removal Cross Section of group $g$}
\nomenclature[G]{$\Sigma_{a,g}$}{Absorption Cross Section of group $g$}
\nomenclature[G]{$\Sigma_{s,g\rightarrow g'}$}{Scattering Cross Section from group $g$ to $g'$}
\nomenclature[G]{$\lambda_{h,i}$}{Decay Heat Constant for group $i$}
\nomenclature[G]{$\bar{\nu}_{g}$}{Average Number of Neutrons produced by fission with energy in group $g$}
\footnotesize{ \printnomenclature }

\appendix

\section{Governing Equations of the Molten Salt Fast Reactor}\label{app: msfr-gov-eqn}

The Molten Salt Fast Reactor \cite{GenIV-RoadMap} is an innovative reactor concept in which the fuel is liquid and homogeneously mixed with the molten salt thermal carrier; this configuration offers significant advantages, for example from the point of view of fuel reprocessing and closure of the fuel cycle, but also poses a lot of unique challenges in terms of monitoring and safety. Moreover, due to the strict coupling between the different physics, even more so than other nuclear reactor configurations, the mathematical modelling of such system requires suitable tools to integrate different physics together, starting from the neutronics and thermal-hydraulics \cite{aufiero2014development}.

The governing equations of the MSFR consist of the Navier-Stokes equations with the energy balance, including turbulence modelling with RANS and the Boussinesq approximation for density variations, the multi-group neutron diffusion and the precursors advection-diffusion: for completeness, this section reports the governing equations and briefly discusses their coupling.

The Navier-Stokes equations for the MSFR reads \cite{versteeg2007introduction}:
\begin{subequations}
\begin{align}
\nabla\cdot \vec{u} =& 0 \label{eq: mass} \\
\frac{\partial\vec{u}}{\partial t} + (\vec{u}\cdot \nabla)\vec{u} =& +\nabla\cdot [\nu_{\text{eff}}(\nabla\vec{u} + (\nabla\vec{u})^{T})] \nonumber \\
&- \frac{1}{\rho} \nabla p + [1 - \beta_{T}(T - T_{0})] \vec{g}  \label{eq: velocity} \\
\frac{\partial T}{\partial t} + \nabla \cdot \left(\vec{u} T\right)  =& \nabla\cdot (\alpha_{\text{eff}}\nabla T) + \frac{q'''}{\rho c_{p}} \label{eq: energy}
\end{align}
\end{subequations}
given $\vec{u}$ as the velocity, $\rho$ as the density, $p$ as the pressure, $\nu_{\text{eff}}$ as the effective kinematic viscosity (accounting for turbulence modelling), $T$ as the temperature, $\beta_T$ as the thermal expansion, $\vec{g}$ as the gravity acceleration, $\alpha_{\text{eff}}$ as the thermal diffusivity, $q'''$ as the power density, which includes both neutron fission and delayed heat sources, and $c_p$ as the specific heat capacity. This set of equations shows the strong coupling between the two different physics; in particular, the source term of the energy equation directly depends on the neutron fluxes, i.e.
\begin{equation}
    q''' = (1-\beta_{h})\sum_{g}\bar{E}_{f,g}\Sigma_{f,g}\phi_{g} + \sum_{i}\lambda_{h,i}d_{i}
    \label{eqn: volumetric-heat-source}
\end{equation}
with $\phi_g$ being the neutron flux of energy group $g$, $\beta_h$ being the fraction of delayed heat source, $\bar{E}_{f,g}$ being the average energy released by fission in group $g$, $\Sigma_{f,g}$ being the fission cross section \cite{DuderstadtHamilton} of group $g$, $\lambda_{h,i}$ being the  decay constant of delayed decay group $d_i$.

Neutronics is governed by the Multi-Group Diffusion Equation \cite{DuderstadtHamilton, aufiero2014development}:
\begin{equation}	
    \frac{1}{v_{g}} \frac{\partial\phi_{g}}{\partial t} - \nabla\cdot (D_{n,g} \nabla \phi_{g}) + \Sigma_{r,g} \phi_{g} = S_{g} 
\label{eqn: diffusion-mg}
\end{equation}
with $D_{n,g}$ being the neutron diffusion coefficient of group $g$ and $\Sigma_{r,g}$ being the removal cross-section accounting for all the reactions which decrease the number of neutrons of group $g$, i.e. absorption $\Sigma_{a,g}$ and infra-group scattering $\Sigma_{s,g \to g'}$:
\begin{equation}
    \Sigma_{r,g}^0 =  \left(\Sigma_{a,g}^0+ \sum_{ g' \neq g }\Sigma_{s,g \to g'}^0 \right)
\end{equation}
where the superscript $0$ refers to the values at the reference temperature. In fact, the cross section are strongly affected by the neutron energy due to the Doppler broadening effect and the variation in density, i.e. 
\begin{equation}
    \Sigma_{r,g} = \left( \Sigma_{r,g}^{0} + A_{r,g}^{0}\ln\frac{T}{T_{0}^{\Sigma}}\right)\cdot \frac{1-\beta_{T}(T - T_{0})}{1 - \beta_{T}(T_{0}^{\Sigma} - T_{0})}
    \label{eqn: xs-mg}
\end{equation}

In the end, the neutron diffusion equations have a source term $S_g$ accounting for the production of neutrons in the group energy $g$ due to prompt fission, decay of precursors and infra-group scattering:
\begin{subequations}
\begin{align}
    S_{g} &= (1-\beta) \chi_{p,g}\,S_{p,g} +\chi_{d,g}\,S_{d} + S_{s,g}\\
      S_{p,g} &= \sum_{g'}\bar{\nu}_{g'}\Sigma_{f,g'}\phi_{g'} \\
      S_{d}   &=\sum_{k}\lambda_{k}c_{k} \\
      S_{s,g} &= \sum_{g' \neq g}\Sigma_{s,g' \to g} \phi_{g'}
\end{align}
\end{subequations}
with $\chi_{p,g}$ and $\chi_{d,g}$ as the fission spectrum of prompt and delayed neutrons and $\bar{\nu}_{g'}$ as the average number of neutrons produced in group $g'$.

When dealing with transient scenarios, the production of neutrons is split into prompt (due to fission) and delayed (due to decay): in the nuclear reactors world, this latter contribution is described by the group precursors $c_k$ which are governed by an advection-diffusion equation in the salt:
\begin{equation}
\begin{split}
\frac{\partial c_{k}}{\partial t} + \nabla\cdot (\vec{u} c_{k}) =& +\nabla\cdot (D_{\text{eff}} \nabla c _{k}) - \lambda_{k} c _{k} \\
&+ \beta_{d,k}\sum_{g} \bar{\nu}_{g} \Sigma_{f,g}\phi_{g}
\end{split}
\label{eqn: precursors}
\end{equation}
with $D_{\text{eff}}$ the effective diffusion coefficient and $\lambda_k$ the decay constant of delayed neutron precursor group $k$. In the end, the precursors contribute also to the energy equation \eqref{eq: energy}, where they are grouped together into the decay heat group $d_i$
\begin{equation}
\begin{split}
    \frac{\partial d_{i}}{\partial t} + \nabla\cdot(\vec{u}d_{i}) =& + \nabla\cdot(D_{\text{eff}}\nabla d_{i}) - \lambda_{h,i}d_{i} \\
    &+ \beta_{h,l}\sum_{g}\bar{E}_{f,g}\Sigma_{f,g}\phi_{g}
\end{split}
\label{eqn: decay-heat-prec}
\end{equation}

\end{document}